\documentclass[fleqn,usenatbib]{mnras}

\usepackage{newtxtext,newtxmath}
\usepackage[T1]{fontenc}

\DeclareRobustCommand{\VAN}[3]{#2}
\let\VANthebibliography\thebibliography
\def\thebibliography{\DeclareRobustCommand{\VAN}[3]{##3}\VANthebibliography}

\usepackage{graphicx}	% Including figure files
\usepackage{amsmath}	% Advanced maths commands
\title[] {Turbulence and the characteristics of circumstellar discs }

\author[R. Riaz,  D.R.G. Schleicher, S. Vanaverbeke, Ralf S. Klessen, and J. Saavedra-Bastidas]{R. Riaz$^{1}$\thanks{E-mail: rafeel.riaz@ubo.cl} 
	D.R.G. Schleicher$^{2}$\thanks{E-mail: dschleicher@astro-udec.cl}  
	S. Vanaverbeke$^{3}$\thanks{E-mail: siegfriedvanaverbeke@gmail.com} Ralf S. Klessen$^{4,5}$\thanks{E-mail: klessen@uni-heidelberg.de} J. Saavedra-Bastidas$^{2}$\thanks{E-mail: jsaavedra2018@udec.cl} \\
	$^{1}$Centro de investigaci\'on en Astronom\'ia, Facultad de Ingeniería, Ciencia y Tecnología, Universidad Bernardo O'Higgins, Av. Viel 1497, Santiago, Chile \\
	$^{2}$Departamento de Astronom\'ia, Facultad Ciencias F\'isicas y Matem\'aticas, Universidad de Concepci\'on, Av. Esteban Iturra s/n Barrio \\
	Universitario, Casilla $160$-C, Concepci\'on, Chile \\
	$^{3}$Centre for mathematical Plasma-Astrophysics, Department of Mathematics, KU Leuven, Celestijnenlaan 200B, 3001 Heverlee, Belgium \\ 
	$^{4}$Universit{\"a}t Heidelberg, Zentrum f{\"u}r Astronomie, Institut  f{\"u}r Theoretische Astrophysik, Albert-Ueberle-Str. 2,  69120 Heidelberg, Germany \\
	$^{5}$Universit{\"a}t Heidelberg, Interdisziplin{\"a}res Zentrum f{\"u}r Wissenschaftliches Rechnen, Im Neuenheimer Feld 205,  69120 Heidelberg, Germany
}

\date{Accepted 2024 May 31. Received 2024 March 31; in original form 2023 March 10}

\pubyear{2024}

\begin{document}
\label{firstpage}
\pagerange{\pageref{firstpage}--\pageref{lastpage}}
\maketitle

% Abstract of the paper
\begin{abstract}
We investigate the properties of circumstellar discs (CDs) produced in hydrodynamical simulations of gravoturbulent core collapse considering Kolmogorov and Burger-type turbulence. We report that massive discs are more prevalent in the Kolmogorov regime than for Burger-type turbulence. A significant number of discs are formed with a radius of $\sim$ 15 au in both cases. However, the number of extended discs with radii $>$ 15 au is significantly larger in case 
of Kolmogorov turbulence. The two regimes of turbulence, in general, yield disc radii in the ranges
of 7 au $-$ 30 au, and 13 au $-$ 39 au, respectively. The corresponding ranges of the disc masses are 30.37 $M_{\rm Jup}$ $-$ 0.92 M$_{\odot}$, and 2.09 $M_{\rm Jup}$ $-$ 0.13 M$_{\odot}$, respectively. Moreover, the ratio $M_{\rm disc}$/$M_{\rm star}$ is higher in models of Kolmogorov-type turbulence than in models of Burgers-type turbulence. We do not find any correlation between $R_{\rm disc}$ and $M_{\rm disc}$ over the explored range of initial temperatures (8 K $-$ 14 K) and the type of turbulence. Also, for these initial thermal variations, the turbulent circumstellar disc structures do not exhibit signs of turbulent diffusion. Nonetheless, both sub and supersonic velocity dispersions cause variations in the specific angular momentum (AM) of infalling gas, especially for CDs with radii $\sim$ 16 au $-$ 21 au. The radial profiles of CDs do not correlate with the initial conditions.
\end{abstract}
% Select between one and six entries from the list of approved keywords.
% Don't make up new ones.

%%%%%%%%%%%%%%%%%%%%%%%%%%%%%%%%%%%%%%%%%%%%%%%%%%

%%%%%%%%%%%%%%%%% BODY OF PAPER %%%%%%%%%%%%%%%%%%
\section{Introduction}
Accretion and transport of angular momentum (AM) are the fundamental processes that play a key role in the formation and evolution of circumstellar discs (CDs) \citep{cassen1981formation, yorke1999formation, machida2016conditions, najita2018protoplanetary}. These fundamental processes include gravitational torques, magnetic fields, hydrodynamic instabilities, and turbulent viscosity which can remove AM. Understanding these fundamental processes in CDs and their relation to the properties of the natal gas core is essential to establishing a plausible connection between the scales of a collapsing molecular gas core and the CDs \citep{jappsen2004protostellar, klessen2010accretion, inutsuka2012present}. This can deepen our knowledge about the physical mechanisms that dictate the accretion process onto the central protostar as well as the formation of planets within its gaseous disc environment \citep{greaves2011all, machida2011origin, seifried2015accretion, kuffmeier2018episodic}. The CDs around young stellar objects (YSOs) also become interesting phenomena when strong molecular outflows from Class 0 protostars(after the collapse in prestellar cores, Class 0 protostars are the first protostellar objects observed
 \citep{andre1993submillimeter}) are taken into consideration \citep{bachiller1996bipolar, beltran2004l1157, tobin2007imaging, downes2007jet, vazzano2021outflows}. \citet{froebrich2006evolution} have defined Class 0 protostars as having masses exceeding $10^{-2}$~M$_\odot$.
 The outflows from Class 0 objects imply that while the matter is still accreting onto the protostar through its CD, it is at the same time  also losing mass via strong jets that generally are produced along the rotational axis of the CD \citep{boss2004molecular}. The turbulence and magnetic fields both affect the formation and evolution of the disc structure around the protostar \citep{hersant2005turbulence, johansen2007rapid, adams2008turbulence, davidson2011magnetic, segura2014magnetic, tsukamoto2016magnetic}. However, there has been a recent debate to determine which one is more influential over the other \citep{wurster2020non}. The fundamental properties of turbulence and magnetism have direct connections to the intrinsic properties of the natal prestellar gas core \citep{evans2003molecular, machida2014conditions}. \citet{pineda2019specific} determined the specific AM profile  $j(r)$ towards two Class 0 objects and a first hydrostatic core candidate in the Perseus cloud. They find a power-law relation that lies in between solid body rotation ($\propto r^{2}$) and pure turbulence ($\propto r^{1.5}$), suggesting that the influence of the initial level of turbulence of the parent gas core is still present at the scale of CDs. Moreover, it is now a well-established fact that a self-gravitating parent core leads to the formation of protostars and that during the early phase of the protostellar evolution (i.e Class 0 phase) protostars are often accompanied by self-gravitating CDs \citep{andre1993submillimeter, kawasaki2021growth}. Nonetheless, the question what factors control the velocity field of the gas around the protostars to transform itself from radial infall to eventual Keplerian rotation needs to be investigated further \citep{brinch2009kinematics, tobin2012complex, yen2013unveiling, aso2015alma, sakai2017vertical}. 
 
 \citet{joos2013influence} have used magnetohydrodynamical models to discuss the role of turbulence-induced misalignment of the CDs which according to their findings favours the massive disc formation around protostars. They explored a turbulent velocity field characterized by a Kolmogorov power spectrum \citep{kolmogorov1941dissipation} for laminar, subsonic, and supersonic initial states of the gas. However, the observations of dense molecular cores of density $10^{-18}$~g cm$^{-3}$ and with sizes of a few
tenths of a parsec show large linewidths that imply supersonic turbulent motions with a power-law spectral velocity distribution consistent with compressible turbulence \citep{zuckerman1974models, heyer2004universality}. This paper investigates the possible effects of the initial thermal state of the prestellar gas core and the nature of its initial turbulent velocity fields i.e Kolmogorov-type (solenoidal) and Burgers-type (compressible) \citep{burgers1948mathematical}) on the formation of disc structures around the protostars. These structures are formed in our hydrodynamical simulation models in which we follow the thermodynamics of the system and the dynamical evolution of the protostars produced during gravoturbulent core collapse.

In a turbulent disc, the momentum of the gas can be diffused
by the turbulent viscosity \citep{takahashi2014two}. Also, \citet{cassen1981formation} have described that the AM of the infalling gas is redistributed by the action of turbulent viscosity on a shear layer near the surface of the disc and the radial shear across cylindrical surfaces parallel to the rotation axis. 

There has been a significant number of recent observational studies performed with a focus on Class 0 and the subsequent phase Class I protostellar disc structures especially with radii > 30 au (see for example \citep{choi2010kinematics, tobin20120, yen2013unveiling, ohashi2014formation, tobin2015sub, lee2016angular, aso2017alma, lee2018alma, tobin2020vla}). Class I objects are sources with mature protoplanetary discs that are still
sustained by a collapsing envelope of material \citep{harsono2014rotationally, aso2015alma}.  Numerical findings in the same context also reproduce discs of similar length scale (see e.g. \citep{xu2021formation}). Nonetheless, it is worth noting that many of the surveys are based on disc sizes in the dust continuum which as they are affected by radial drift may not be directly compared to the sizes of gas discs. For instance, \citet{maret2020searching} used Keplerian motion as a criterion for detection down to 50 au, but cannot also strongly rule out the presence of larger discs given the ambiguity introduced by optical depth effects inherent in embedded systems. With this caveat, we present a comparison between numerical findings and observational claims where for the simulations the disc radii are being measured based on a particle membership criterion that does not directly translate to the detection limits of the observations. We report results related to the protostellar disc structures that appear in our simulations and we do not find Class 0 discs with radii > 40 au.

Moreover \citet{stamatellos2011importance} have modelled the episodic accretion process on a protostar which takes place mainly by accretion from the surrounding disc. for their initial condition with a cloud core mass of 5.4 M$_{\odot}$ and core size 50,000 au, the core collapses to form a disc structure around a primary protostar with a disc radius of $\geq$ 50 au.
This result provides insights into the dynamics and characteristics of protostellar formation, particularly the role of episodic accretion and the dominance of accretion from the disc during this process. \citet{allen2003collapse} proposed that, in the absence of angular momentum transport, one would expect the formation of a rotationally supported disc with an approximate size of around 33 au. However, observations of a low-luminosity protostar in the nearby Taurus molecular cloud (IRAM 04191) indicated a disc size limit of less than 10 au. This observational limit suggests that the protostellar disc around IRAM 04191 is relatively small. They attributed the small disc size to efficient magnetic braking, which prevents the formation of a more massive or larger disc in close proximity to the star. This efficient magnetic braking mechanism appears to work against the idea of significant mass accumulation in a disc around the protostar. Also, according to \citet{price2007impact}, for intermediate magnetic field strengths, there is a discernible trend in both the timing of protostar formation and the subsequent size of the formed disc. This trend suggests that as the magnetic field strength increases, the formation of the protostar takes place progressively later. Additionally, the size of the formed disc within the range of 30 au to 50 au decreases as the magnetic field strength is increased (see Fig. 3 and 4 in their work). This implies that the magnetic field strength plays a significant role in influencing the temporal evolution of protostar formation and the size of the resulting protostellar disc. \citet{busquet2019unveiling} performed an observational study of the disc-star systems. They focused on a cluster of 25 continuum sources, with the suggestion that these sources likely trace discs around Class 0/I protostars. According to the results, the median value of the disc radius in this cluster is reported to be 34 au. This disc radius is noted to be smaller than the median disc radius observed in the Taurus molecular cloud (which is 92 au), but it is comparable to the values found in other regions such as Ophiuchus and the Orion Nebula Cluster. These findings suggest some variations in disc sizes among different star-forming regions. In addition to \citet{busquet2019unveiling}, \citet{tobin2020vla} have conducted a survey of 328 protostars in the Orion molecular clouds with the Atacama Large Millimeter/ submillimeter Array at 0.87 mm. From their results, it can be concluded that protostellar (and specifically Class 0) discs larger than 50 au are not rare.
 %The numerical finding reported in this paper is therefore consistent with CALYPSO IRAM-PdBI survey which suggests that large Class 0 discs are rare \citep{hennebelle2016magnetically, maury2019characterizing}.  

The structure of this paper is summarized as follows.  We describe our method along with the initial conditions adopted for our simulations in section 2. In section 3, we present our results and discuss them in detail. Finally, in section 4, we provide the conclusions and outlook. 
%\begin{figure*}
	 %To include a figure from a file named example.*
	% Allowable file formats are eps or ps if compiling using latex
	% or pdf, png, jpg if compiling using pdflatex
%	\includegraphics[angle=0,scale=0.432]{FIGURE2.png}
%	\caption{Global disc morphology for models M1a$-$M4a (top panels) and M1b$-$M4b (bottom panels) at the end of our computation when star formation efficiency (SFE) in each model reaches $\xi$ = 2 \%. In each row from left to right, the initial turbulent velocity field corresponding to $\mathcal{M}$ = 0.1 $-$ 0.8 is imposed onto a solid-body rotation. Each panel (the xz-plane) shows in logarithmic scale the projected column density ($ \Sigma$) integrated along the y-axis in g cm$^{-2}$. Sink particles (protostars) are shown as red dots. Colour in the online edition.}
%	\label{fig:figur6}
%\end{figure*}

\section{Methods}
The simulations discussed in this paper were originally published in \citet{riaz2021turbulence} hereafter RSVK explaining the phenomenon of episodic accretion. We now use the same calculations to address characteristics of CDs that are found associated with most of the protostars which are formed in our models. In section 2.1 we provide a summary of the method used to perform the simulations, which is then followed by section 2.2 where we summarise the initial conditions that also help to comprehend the results presented in this paper. Also for the present work, section 2.2 presents a detailed description of the method we adopted to analyse the disc structures around protostars which are produced during the gravoturbulent collapse of Jeans unstable molecular gas cores.
\vspace{-0.25cm}
\subsection{The gravoturbulent core collapse calculations}
Referring to our recent work \citep{riaz2021turbulence}, we performed simulations using the particle method known as smoothed particle hydrodynamics (SPH). We utilised the computer code GRADSPH \footnote{Webpage GRADSPH: http://www.swmath.org/software/1046} which has been developed by \citet{vanaverbeke2009gradsph}.
Previous studies point to the chaotic nature of the physical processes involved in star-formation, which is typically followed by the formation of CDs \citep{klein2001fragmentation,jappsen2005stellar,goodwin2004simulating,takaishi2020star,wollenberg2020formation,guszejnov2022effects,concha2023evolution}. This chaotic nature of star formation stems from the high sensitivity to initial conditions, resulting in diverse outcomes even with minor variations in parameters like the initial mesh size, the positioning of SPH particles or turbulence seeds. Consequently, studying and establishing general trends regarding the impact of factors like the initial level of turbulence or the turbulence spectral index first on star formation and second on formation of disc structure around these newly formed stars requires a broader approach. Instead of relying solely on individual simulations or single realizations, considering a range of runs with diverse initial conditions becomes crucial. This approach helps in deriving statistical insights and identifying overarching patterns amidst the inherent variability. By analyzing a sample of runs with varied initial turbulence seeds, we can better understand the broader spectrum of outcomes and extract more reliable generalizations about the effects of turbulence on the circumstellar disc formation and evolution. While Burgers turbulence is strongly motivated by observations of supersonic turbulence in molecular clouds \citep{hennebelle2012turbulent,boneberg2015turbulence}, which causes shocks and was suggested to give rise to a steeper turbulent spectrum, the standard turbulence scenario is based on the Kolmogorov scenario, which predominantly considers the decay of turbulent eddies  \citep[see also][]{mac2004control}.

For turbulence realisation, we consider two different types of turbulent spectra in our initial conditions, with slopes as appropriate for Kolmogorov \citep{kolmogorov1941dissipation} and Burgers type \citep{burgers1948mathematical} turbulence. The former is incompressible, subsonic turbulence while the latter refers to supersonic, shock-dominated turbulence that can promote formation of dense structures inside the collapsing gas. We inject a spectrum of turbulence into the initial conditions by adding the superposition of the velocity of 1000 shear waves (transverse waves) with random propagating directions to the initial velocity of each particle. The wavelength $\lambda$ of the shear waves is distributed uniformly between $0.001 R$ and $R$, where $R$ is the radius of the gas core, while the amplitude $A$ of the waves follows a spectrum with $A \sim \lambda^{p}$, with the index $p$ taking values of 5/3 and 2 in our models for the Kolmogorov and Burgers type turbulence, respectively. The amplitude of the resulting turbulent velocity field is then rescaled so that the RMS Mach number of the turbulent flow with respect to the initial isothermal sound speed equals the value in each model. After the generation of the initial conditions, there is no driving of turbulence, but it may dynamically develop, including decay but also driving by infall. In both cases, we consider Mach numbers $\mathcal{M}$ = 0.75 and $\mathcal{M}$ = 3.45, respectively (see Table 1).
We used sink particles to represent protostars in our simulations. We set a constant accretion radius $r_{\rm acc}$ = 1 au for the sink particles, which always remains greater than the Jeans length during the gas collapse. Whenever the density during the gas collapse reached a specific value of $10^{-11}$~g cm$^{-3}$, a sink particle was introduced inside the gas to avoid the “Courant catastrophe" (see e.g. \citep{bate1995modelling, bromm2004accretion, federrath2010modeling}).

During the gravoturbulent collapse of the molecular core the gas attained the value $10^{-11}$~g cm$^{-3}$ at various locations inside the densest part of the collapsed regions. Thus all the sink particles (i.e. protostars) in this work are numbered by following the order in which they are formed during the evolution of each model. The method described by \citet{stacy2013constraining} was used to treat mergers of sink particles in our simulations when the following three criteria were satisfied: 
	\begin{itemize}\setlength{\itemsep}{0.25cm}
		\item When their relative distance $d$ is smaller than $r_{\rm acc}$ (1 au) so that $d < r_{\rm acc}$.
		\item When the total energy $E_{\rm tot}$ of the pair of sink particles is negative so that the pair is gravitationally bound.
		\item When the least massive sink (secondary) of the pair has insufficient angular momentum to remain rotationally supported against infall onto the massive sink (primary) i.e.  $j_{\rm sec}$ $<$ $j_{\rm cent}$, where $j_{\rm cent}~$=$~\sqrt{G~M_{\rm primary}~d}$ and $M_{\rm primary}$ denotes the mass of the most massive sink of the pair. 
	\end{itemize} 
 A schematic which illustrates the criterion followed to determine the disc membership based on the chosen value of the density threshold is provided in Figure 1.
\vspace{-0.5cm}
\subsection{How we treat binary and isolated sinks}	
In order to determine pairs of sinks which form gravitationally bound binary systems, we define the total orbital energy per unit mass of a pair of sink particles as in \cite{stacy2013constraining}:
\begin{equation} \label{orbitalenergy}
\epsilon=\epsilon_{p}+\epsilon_{k},
\end{equation}
where $\epsilon_{p}$ and $\epsilon_{k}$ are the gravitational potential energy and kinetic energy per unit mass, respectively, and are defined as follows
\begin{equation} \label{potentialenergy}
\epsilon_{p}= -\frac{G\left(M_{1}+M_{2}\right)}{r},
\end{equation}
and
\begin{equation} \label{kineticenergy}
\epsilon_{k}=\frac{1}{2}v^{2},
\end{equation}

knowing that $r$ is their mutual distance, $v$ is their relative velocity, and $M_{1}$ and $M_{2}$ are the masses of the pair, respectively. A pair of sinks is considered a binary if $\epsilon < 0$. Sink particles which do not fulfil these criteria are treated as isolated protostars.
%Given the masses, the positions and the velocities of sink particles which are components of a binary, we determine the instantaneous orbital elements of the binary using standard formulae from celestial mechanics. We refer the reader to chapter 6 of \citet{Danby} for details.
\vspace{-0.25cm}

\begin{figure}
	 %To include a figure from a file named example.*
	% Allowable file formats are eps or ps if compiling using latex
	% or pdf, png, jpg if compiling using pdflatex
	\includegraphics[width=\columnwidth]{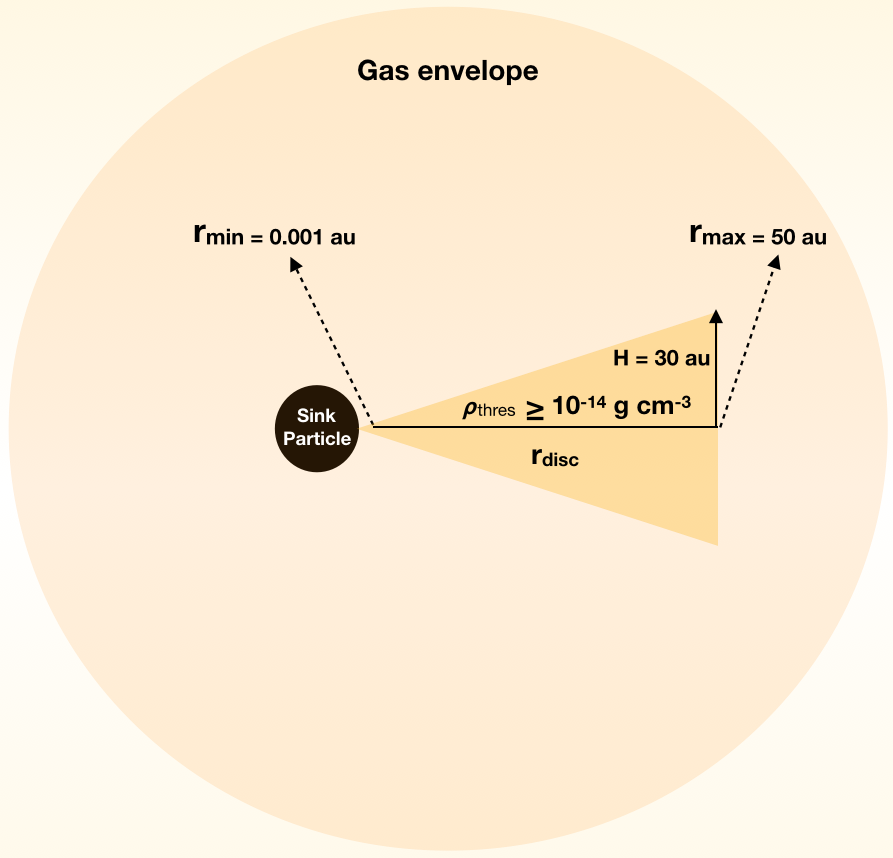}
	\caption{A schematic illustration for the criterion followed to determine the disc membership based on the chosen value of density threshold i.e. $\rho_{\rm thres}$ $\geq$ $10^{-14}$~ g cm$^{-3}$ is shown.}
	\label{fig:figur6}
\end{figure}
\subsection{Method to explore disc properties}
We now describe the method we follow to define the scale of the CD for each sink particle (protostars) that is formed in our simulation models. Our method to identify disc takes inspiration from the work performed by \citet{joos2012protostellar, gray2018effect, hirano2020effect}. Throughout our calculations for characterising the disc structure in this paper, we set the threshold density $\rho_{\rm thres} = 1.0 \times 10^{-14}$~ g cm$^{-3}$ to determine the protostellar disc radius $r_{\rm disc}$. Any SPH particle in the vicinity of the protostar that has a density $\rho_{\rm i}$ $\geq$ $\rho_{\rm thres}$ may qualify for the disc membership (see figure 1). While approximating the disc dimension associated with the protostars we set the inner radius of the disc $r_{\rm min}$, the outer radius of the disc $r_{\rm max}$, and the scale height $H$ of the disc by taking into account the appearance of discs in the collapsed state of the parent gas core. If the protostars exist in close proximity then both ($r_{\rm min}$, $r_{\rm max}$) and $H$ are adjusted accordingly to avoid the overlapping of the CD structures. In addition to the density criterion $\rho_{\rm i}$ $\geq$ $\rho_{\rm thres}$ which sets the volume $V$ of SPH particles, we perform the following tests for the disc membership: 
	\begin{itemize}\setlength{\itemsep}{0.25cm}
		\item test (1): Within the volume $V$ that potentially defines the disc material we check for all sph particles whose azimuthal velocities $v_{\rm \phi}$ are larger than a few times the infall (radial) velocities $v_{\rm r}$ 
		
		i.e. $|v_{\rm \phi}|$ > $f_{\rm thres}$.$|v_{\rm r}|$, where $f_{\rm thres}$ = 2.0
		\item test (2): For the expected hydrostatic equilibrium of the disc, the azimuthal velocity $v_{\rm \phi}$ should be  larger than the vertical velocity ($v_{z}$)
		i.e. $|v_{\rm \phi}|$ > $f_{\rm thres}$.$|v_{z}|$
		\item test (3): The rotational support ($\rho_{\rm i}$ $v_{\rm \phi}^{2}$ / {2}) of the disc sholud also be larger than the thermal support
		i.e. $\rho_{\rm i}$ $v_{\rm \phi}^{2}$ / {2} $>$ $p_{i}$, where $p_{i}$ = ($\gamma$ - 1) $u_{\rm i}$ $\rho_{\rm i}$, and where $u_{\rm i}$ is the internal energy of the gas.
	\end{itemize}
%We approximate the disc-core inclination $\theta_{\rm cd}$ in the xz-plane of the collapsed gas core by visually inspecting each candidate disc fulfilling the above set of criteria. This ensures that each disc segment lies on a common equatorial plane constituting the semi-major axis of the CD. In the case of a circular disc morphology, we declare the disc orientation as unknown in the absence of semi-major axis.

%\citet{yen2015observations} have argued that 
%Analyse the presence of Keplerian disc and the degree of misalignment of the CD with axis of rotation of the parent gas core (Ref... doi 10.1088/0004-637X/799/2/193). 
\citet{elsender2021statistical} have suggested that if the accretion radius for sink particle is set to ${r_{\rm acc}}$ = 1 au then associated CD can only be resolved if it has a radius ${R_{\rm disc}}$ $\geq$ 1 au. In our calculations, we also set ${r_{\rm acc}}$ = 1 au and hence do not spatially resolve any CD that has a ${R_{\rm disc}}$ < 1 au.

\begin{table}
	\centering
	\caption{Summary of the initial physical parameters of the simulation models M1a$-$M8a and M1b$-$M8b \citep{riaz2021turbulence}. The table describes the initial gas temperature ($T_{\rm i}$), the density at which the transition from isothermal to adiabatic collapse occurs ($\rho_{\rm crit}$), the turbulent Mach number ($\mathcal{M}$), the ratio of thermal energy to the gravitational potential energy of the gas core ($\alpha_{\rm th}$), the ratio of turbulent kinetic energy to the gravitational potential energy of the gas core ($\alpha_{\rm turb}$), and the energy spectral index $p$. For each model, the total mass inside the core, the initial radius of the core, and the initial uniform gas density ($\rho_{\rm i}$) of the gas core are $5.0$~M$_{\odot}$, $0.027$~ pc, $3.845 \times 10^{-18}$~ g cm$^{-3}$, respectively.} 
	\label{tab:Table1}
	\begin{tabular}{ccccccc} % four columns, alignment for each
		\hline
		\hline
		Model &$T_{\rm i}$ (K)&$\rho_{\rm crit}$~(g cm$^{-3})$& $\mathcal{M}$ & $\alpha_{\rm th}$& $\alpha_{\rm turb}$& index $p$ \\
		\hline
		M1a & 8 &$1.1 \times 10^{-13}$& 0.75 & 0.090& 0.017 & 5/3\\
		M2a & 10 &$1.9 \times 10^{-13}$& 0.75 & 0.113& 0.021& 5/3\\
		M3a & 12 &$3.0 \times 10^{-13}$& 0.75 &0.136 & 0.025& 5/3\\
		M4a & 14 &$4.5 \times 10^{-13}$& 0.75 & 0.159& 0.029& 5/3\\
		M5a & 8 &$1.1 \times 10^{-13}$& 3.45 & 0.090& 0.360& 2\\
		M6a & 10 &$1.9 \times 10^{-13}$& 3.45 & 0.113& 0.450& 2\\
		M7a & 12 &$3.0 \times 10^{-13}$& 3.45 & 0.136& 0.541& 2\\
		M8a & 14 &$4.5 \times 10^{-13}$& 3.45 & 0.159& 0.631& 2\\
		M1b & 8 &$1.1 \times 10^{-13}$& 0.75 & 0.090& 0.170& 5/3\\
		M2b & 10 &$1.9 \times 10^{-13}$& 0.75 & 0.113& 0.021& 5/3\\
		M3b & 12 &$3.0 \times 10^{-13}$& 0.75 & 0.136& 0.025& 5/3\\
		M4b & 14 &$4.5 \times 10^{-13}$& 0.75 & 0.159& 0.029& 5/3\\
		M5b & 8 &$1.1 \times 10^{-13}$& 3.45 & 0.090& 0.360& 2\\
		M6b & 10 &$1.9 \times 10^{-13}$& 3.45 & 0.113& 0.450& 2\\
		M7b & 12 &$3.0 \times 10^{-13}$& 3.45 & 0.136& 0.541& 2\\
		M8b & 14 &$4.5 \times 10^{-13}$& 3.45 & 0.159& 0.631& 2\\
		\hline
	\end{tabular}
\end{table}
\vspace{-0.25cm}
\subsection{The initial conditions}
In our simulations presented in RSVK, we used 250025 SPH particles in each model to construct the gas core. For every SPH particle the number of neighbouring particles was set to $N_{\rm opt}$ = 50. Thus, following the criterion $M_{\rm resolution}$ = 2 $N_{\rm opt}$ $m_{\rm particle}$, we had in our simulations a minimum resolvable mass $M_{\rm resolvable} = 1.999 \times 10^{-3}$~M$_\odot$. To show that the circumstellar disc structures, which form in our simulations, follow the required mass resolution criterion, we provide in Appendix A a resolution study to support the disc analysis presented in this paper.

A total of sixteen models which were divided into two main sets M1a$-$M8a and M1b$-$M8b were run where each set represented a different initial seed that was used to generate the initial turbulent velocity structure inside the spherical gas core (see RSVK for more details). The total mass of the gas core was 5 M$_{\odot}$ and the radius is 0.027 pc.  The gas in every model had a uniform initial gas density $\rho_{\rm i} = 3.8 \times 10^{-18}$~ g cm$^{-3}$ which was consistent with the IRAS Sky Survey of bright cores (L1544 and L1689B) from the region of Taurus–Auriga \citep{kirk2005initial}. Inspired by previous observational findings related to the thermal state of prestellar gas cores (see e.g. \citep{pagani2007depletion, bergin2006thermal, bacmann2016origin}) we set four different values (8, 10, 12, and 14 K) as the initial temperatures of the gas cores.
The free fall time $t_{\rm ff}$ = $\sqrt{\frac{3 \pi}{32 G \rho_{\rm i}}}$ and the ratio of the rotational energy to the gravitational potential energy of the core $\beta$ = $\frac{R^{3}\omega^{2}}{3 G M}$ were set to 30.627 kyr and 0.0785, respectively. The other important ratios such as the ratios of the thermal and turbulent energies ($U_{\rm th}$, $U_{\rm turb}$) to the gravitational potential energy ($\Omega$) were described with the parameters $\alpha_{\rm th}$ = $\frac{5 R k T}{2 G M \mu m_{\rm H}}$ and $\alpha_{\rm turb}$ = $\frac{U_{\rm turb}}{|\Omega|}$ remained model dependent (see also Table 1 for the values of involved quantities). 

Two different types of turbulent spectra were considered in our initial conditions, with slopes as appropriate for Kolmogorov-type \citep{kolmogorov1941dissipation} and Burgers type \citep{burgers1948mathematical} turbulence. The former was incompressible, subsonic turbulence modelled with Mach numbers $\mathcal{M}$ = 0.75, while the latter referred to supersonic, shock-dominated turbulence, which was modelled with $\mathcal{M}$ = 3.45 (see also Table 1). 

We used the following barotropic equation of state (EOS) to capture the thermodynamical evolution during gravoturbulent collapse of the molecular gas core:
\begin{equation} \label{EOS}
P=\rho c_{s}^{2}\left[1+\left(\frac{\rho}{\rho_{\rm crit}}\right)^{\gamma-1}\right],
\end{equation}
where $c_{s}$ is the sound speed, $\gamma$ = 5/3 is the adiabatic index, and the critical density \textit{$\rho$}$_{\rm crit}$ (which is a model-dependent parameter in our calculations, see Table 1) refers to the phase transition from isothermal to the adiabatic collapse of the gas core. \citet{omukai2005thermal} have taken into account the balance between cooling which is dominated by continuum emission via thermal radiation from the dust and compressional heating. When the gas becomes adiabatic, the resulting relation between the transition temperature and the number density of the gas is  
\begin{equation} \label{transition}
T=\left(\frac{k^{3}}{12\sigma^{2}m_{\rm H}}\right) ^{1/5} {n^{2/5}_{\rm H}},
\end{equation}
where $k$, $\sigma$, $m_{\rm H}$, and $n_{\rm H}$ are the Boltzmann constant, Stefan-Boltzmann constant, mass of the hydrogen atom, and the number density of the gas, respectively. We treat the gas as fully molecular. Since the collapsing gas in our models remains initially in the isothermal phase, the value of \textit{$\rho$}$_{\rm crit}$ which marks the phase transition from isothermal to adiabatic collapse is obtained from equation 5.

\begin{figure}
	 %To include a figure from a file named example.*
	% Allowable file formats are eps or ps if compiling using latex
	% or pdf, png, jpg if compiling using pdflatex
	\includegraphics[width=\columnwidth]{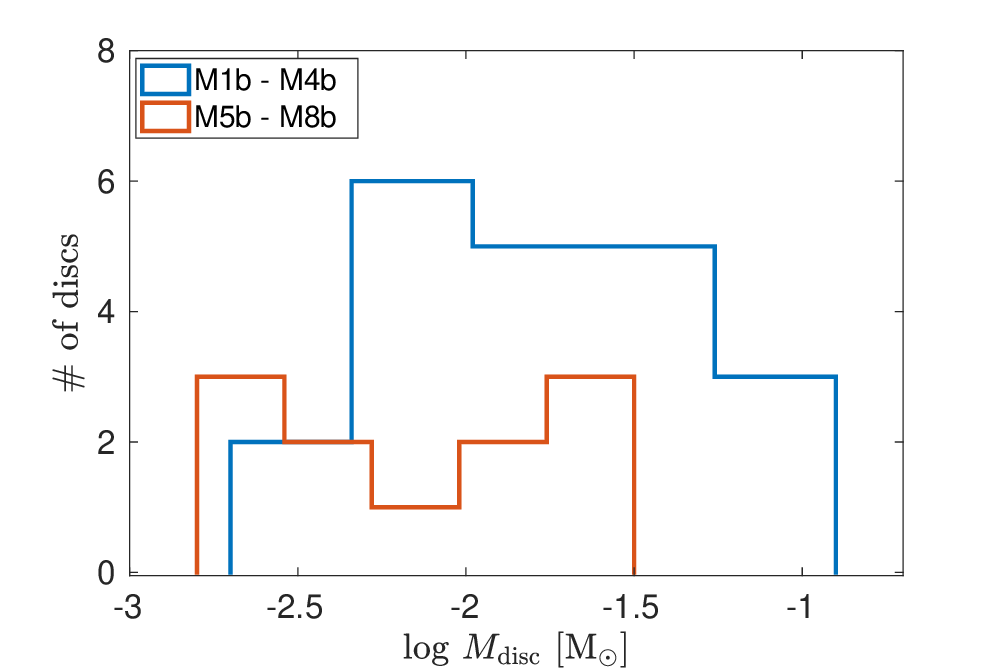}
	\includegraphics[width=\columnwidth]{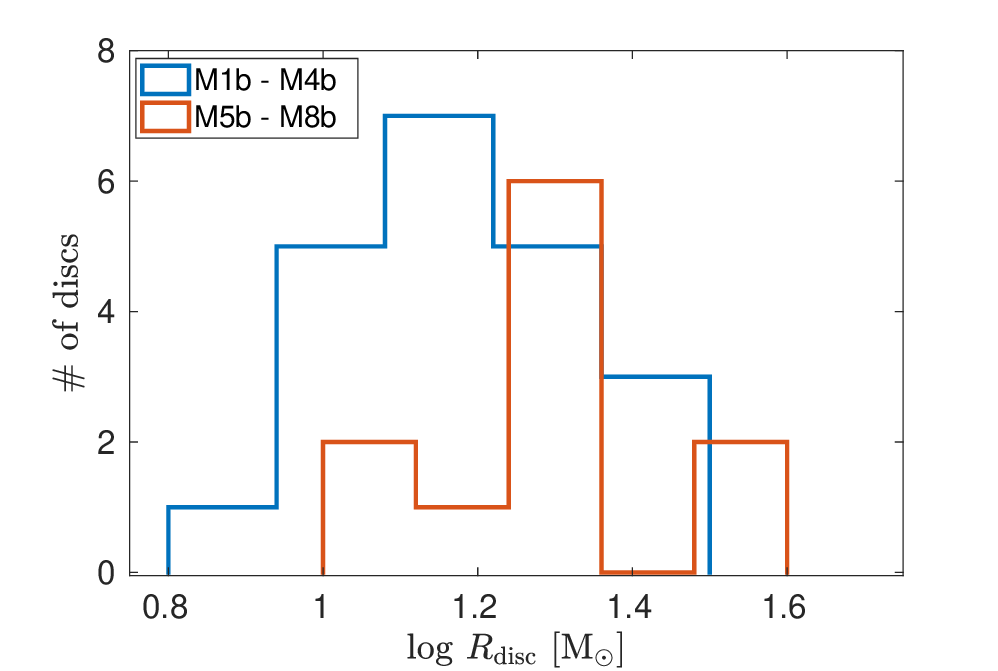}
	\caption{Distributions of disc mass (top panel) and disc radius (bottom panel) for models M1b$-$M8b at the end of our computation when star formation efficiency (SFE) in each model reaches $\xi$ = 15 \%, except models M7b and M8b where simulations due to high computational cost are terminated at $\xi$ = 10 \%. The two types of models of turbulence are presented in the distributions of disc mass and radius without taking into account the initial thermal states of the collapsing molecular gas cores as well as with no classification of the isolated and binary configurations. The disc radius and mass are given in units of au and solar mass, respectively. The solid blue and red lines indicate mass of the disc in model sets M1b$-$M4b and M5b$-$M8, respectively. Colour in the online edition.}
	\label{fig:figur6}
\end{figure}

\begin{figure}
	 %To include a figure from a file named example.*
	% Allowable file formats are eps or ps if compiling using latex
	% or pdf, png, jpg if compiling using pdflatex
	\includegraphics[width=\columnwidth]{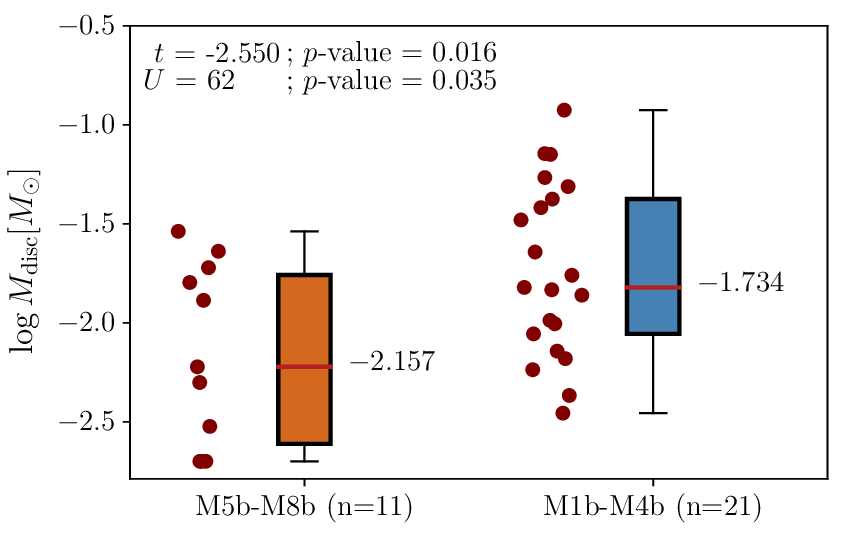}
	\includegraphics[width=\columnwidth]{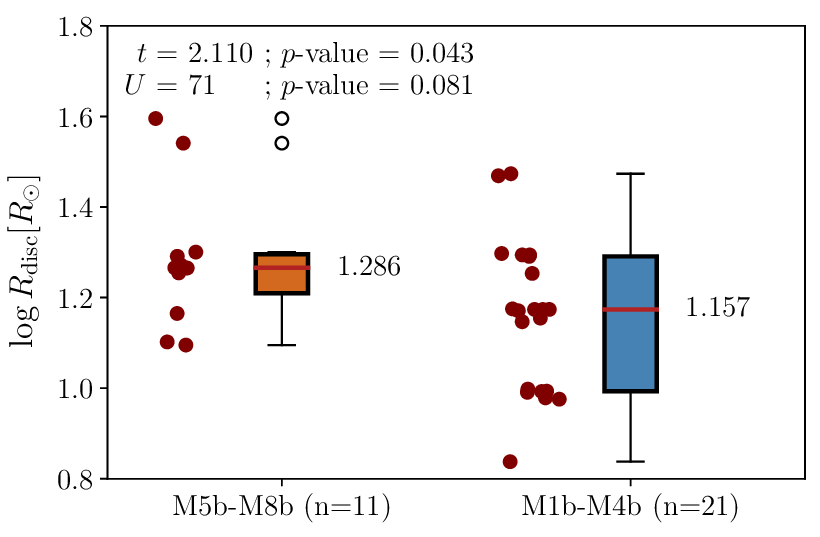}
	\caption{Box-plots with the distribution of masses (top panel) and radii (bottom panel) of protoplanetary discs with Kolmogorov-type and Burger-type turbulence. The boxes represent the interquartile range (IQR) measured from the 25 to the 75 percentile. The whiskers represent the maximum an minimum values expected around the IQR. We highlight the number of samples per distribution, include the $t$ statistic and $p$-value obtained in the $t$-test, the $U$ statistic and $p$-value obtained in the $U$-test between both distributions and their respective mean. The red dots are the actual sample values. Colour in the online edition.}
	\label{fig:figur6}
\end{figure}

\begin{figure}
	 %To include a figure from a file named example.*
	% Allowable file formats are eps or ps if compiling using latex
	% or pdf, png, jpg if compiling using pdflatex
%	\includegraphics[width=\columnwidth]{myradiusmass1.png}
%	\includegraphics[width=\columnwidth]{myradiusmass2.png}
 	\includegraphics[width=\columnwidth]{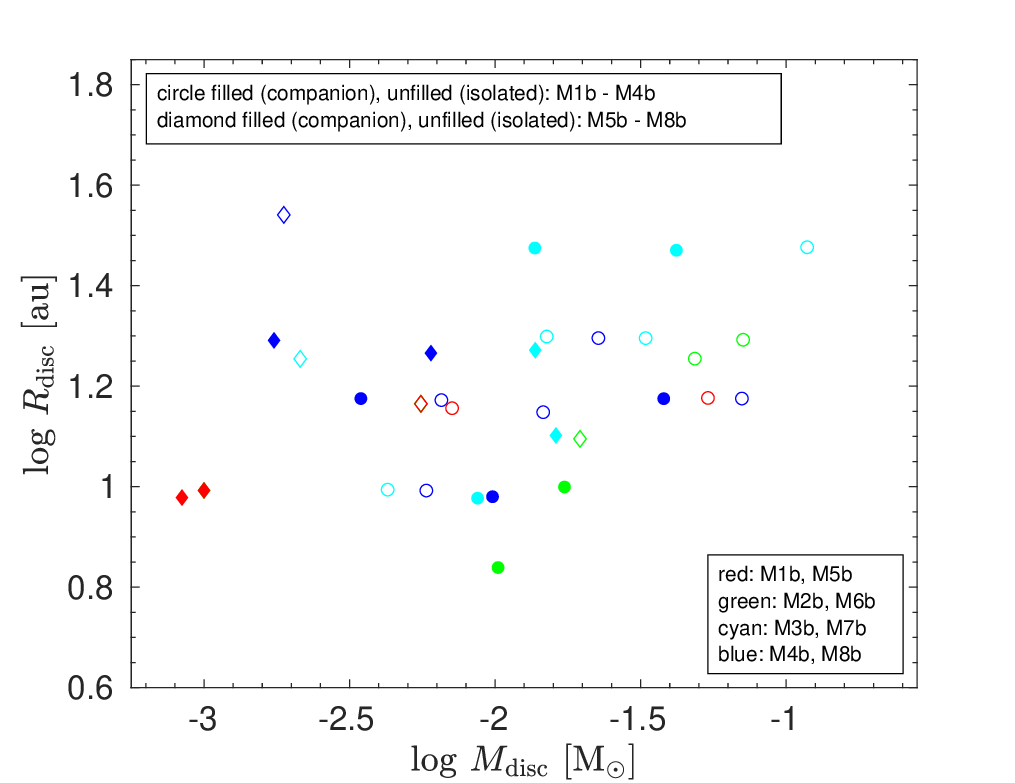}
	\caption{Disc radius as a function of disc mass for models M1b$-$M8b at the end of our computation when star formation efficiency (SFE) in each model reaches $\xi$ = 15 \%, except models M7b and M8b where simulations due to high computational cost are terminated at $\xi$ = 10 \%. The disc radius and mass are given in units of au and solar mass, respectively. Filled and unfilled circles indicate the discs which are part of the binary systems and the isolated systems, respectively in model set M1b$-$M4b. Similarly, filled and unfilled diamonds indicate the protostars which are part of the binary systems and the isolated systems, respectively in model set M5b$-$M8b. Colour red, green, cyan, and blue represent models M1b and M5b, M2b and M6b, M3b and M7b, and M4b, and M8b, respectively. Colour in the online edition.}
	\label{fig:figur6}
\end{figure}

%\begin{figure*}
	 %To include a figure from a file named example.*
	% Allowable file formats are eps or ps if compiling using latex
	% or pdf, png, jpg if compiling using pdflatex
%	\includegraphics[angle=0,scale=0.5675]{sigma_seed1.png}
%	\caption{Simulation results for models M1a$-$M4a (top panels) and M1b$-$M4b (bottom panels) at the end of our computation when star formation efficiency (SFE) in each model reaches $\xi$ = 2 \%. Each panel (the xy-plane) shows in logarithmic scale the projected column density ($ \Sigma$) integrated along the z-axis in g cm$^{-2}$. Sink particles (protostars) are shown as red dots. The arrows in each panel mark the primary (p) and secondary (s) components of MMPB located in the cluster. Colour in the online edition.}
%	\label{fig:figur6}
%\end{figure*}

\begin{figure}
	 %To include a figure from a file named example.*
	% Allowable file formats are eps or ps if compiling using latex
	% or pdf, png, jpg if compiling using pdflatex
     \includegraphics[width=\columnwidth]
     {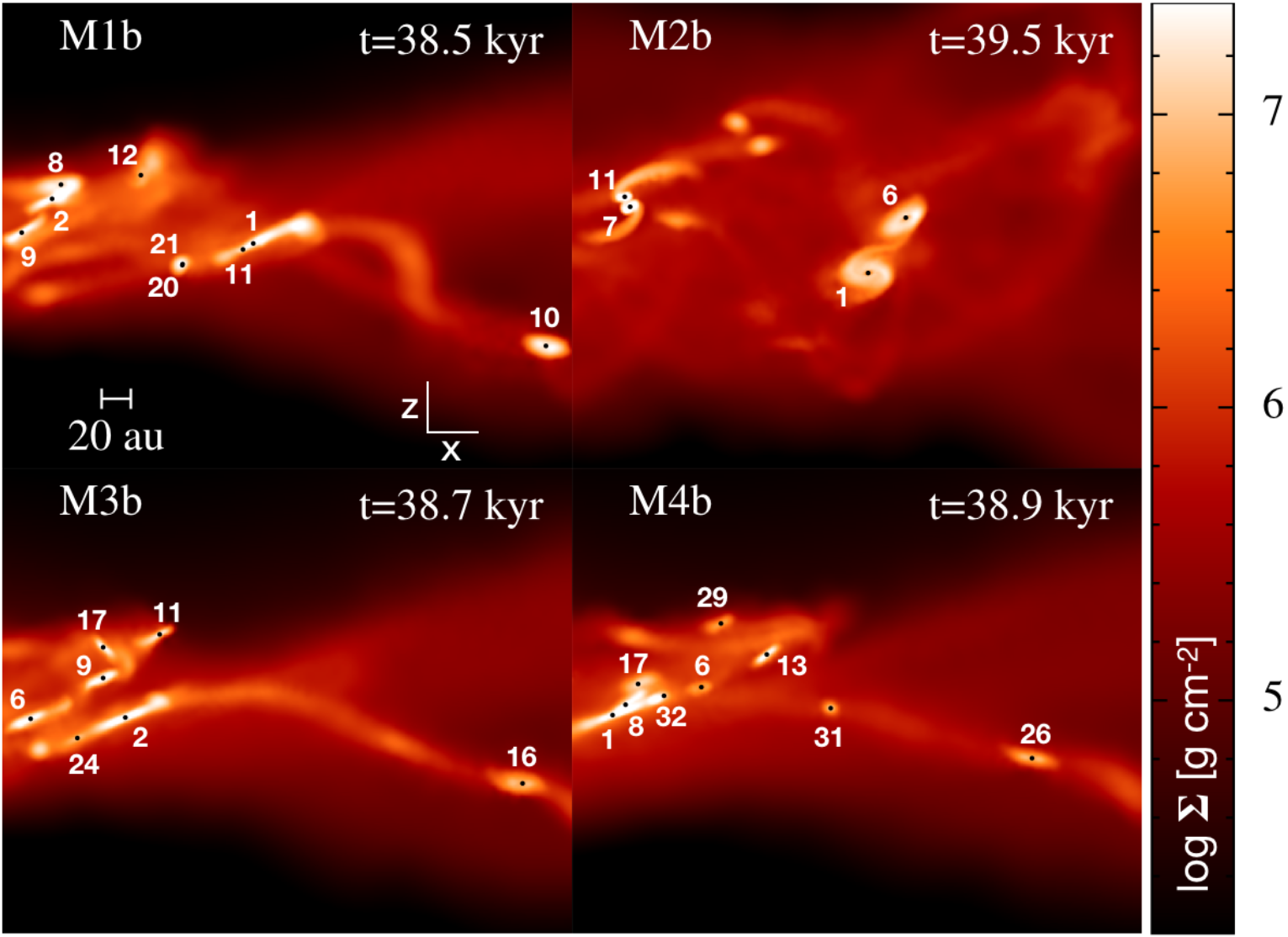}
	\caption{Morphology of the natal core collapse for models M1b$-$M4b at the end of our computation when star formation efficiency (SFE) in each model reaches $\xi$ = 15 \%. Each panel (the xz-plane) shows in logarithmic scale the projected column density ($ \Sigma$) integrated along the y-axis in g cm$^{-2}$. Sink particles (protostars) are shown as black dots and are all numbered by following the order in which they are formed during the evolution of each model. The spatial scale shown in the top-left panel is also relevant for rest of the panels in the figure. Each panel represents a box size of 300 x 400 au. Colour in the online edition.}
	\label{fig:figur6}
\end{figure}

\begin{figure}
	 %To include a figure from a file named example.*
	% Allowable file formats are eps or ps if compiling using latex
	% or pdf, png, jpg if compiling using pdflatex
 \includegraphics[width=\columnwidth]
 {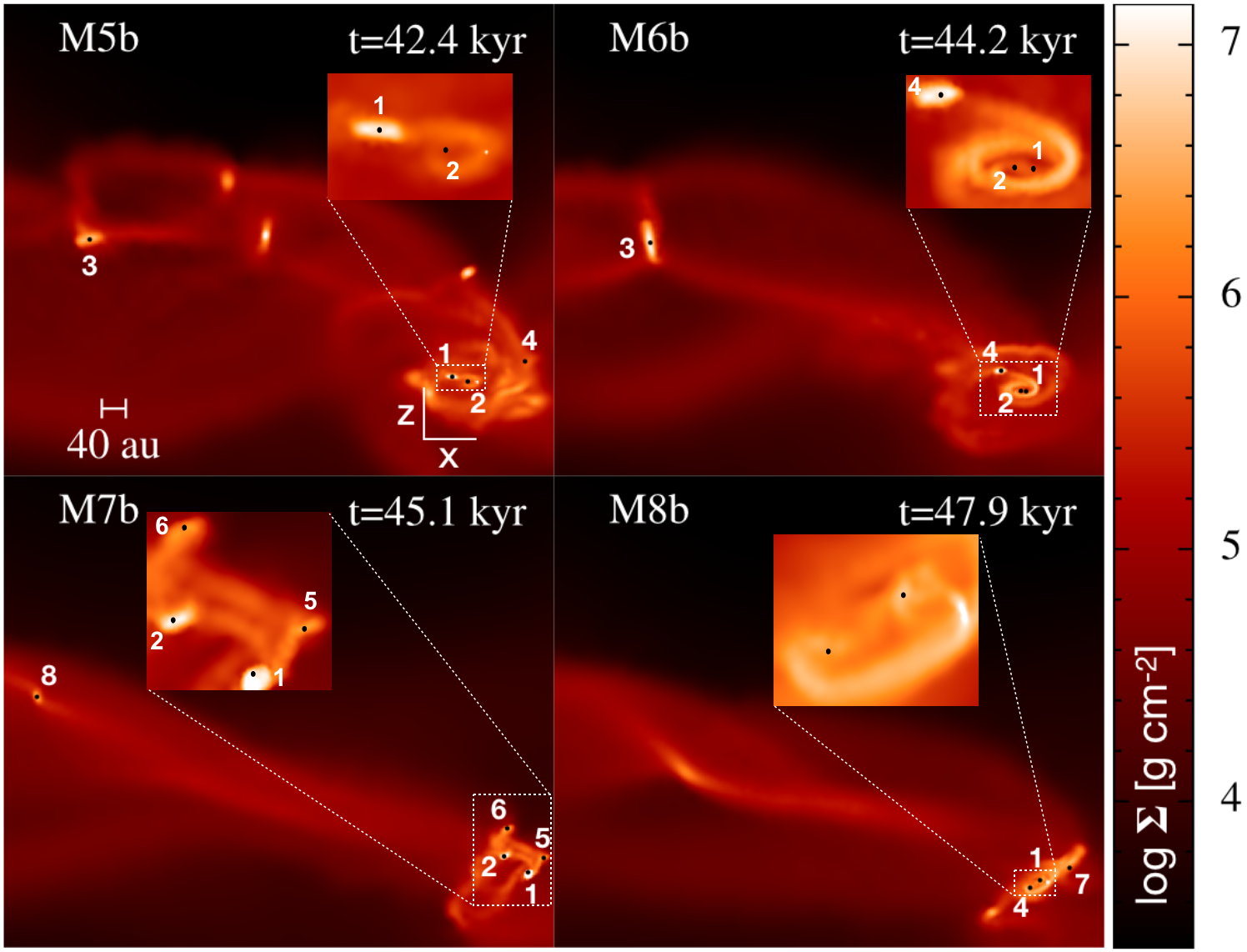}
	\caption{Morphology of the natal core collapse for models M5b$-$M8b at the end of our computation when star formation efficiency (SFE) in models M5b and M6b reaches $\xi$ = 15 \%, and in models M7b and M8b reaches $\xi$ = 10 \%. Each panel (the xz-plane) shows in logarithmic scale the projected column density ($ \Sigma$) integrated along the y-axis in g cm$^{-2}$. Sink particles (protostars) are shown as black dots and are all numbered by following the order in which they are formed during the evolution of each model. The spatial scale shown in the top-left panel is also relevant for rest of the panels in the figure. Each panel represents a box size of 880 x 1000 au. Colour in the online edition.}
	\label{fig:figur6}
\end{figure}

\begin{table} \label{tbl-1}
%\begin{flushleft}
\centering
\caption{Summary of the models in set M1b$-$M4b with the second random seed (corresponding to two different realisations of turbulence, with the same statistical properties). The table is constructed for sink particles (protostars) that form around them a CD structure. Only Class 0 objects are included in the table. The entire table is constructed at the points in time when SFE $\xi$ reaches 15 \% in each model. The table describes the identity of protostar (Sink), the mass of the protostar ($M_{\rm star}$), the mass of the disc ($M_{\rm disc}$), the radius of the disc ($R_{\rm disc}$), and the disc$-$star mass ratio ($M_{\rm disc}$/$M_{\rm star}$). The mass and the radius are provided in solar mass units and in au, respectively.    }
%\headline{$\xi$ = 2 \%}

%. $\parallel$. $\perp$ symbols for parallel and perpendicular
\label{tab:Table1}
\begin{tabular}{ccccc} % four columns, alignment for each

%\begin{tabular}{ccccc}
\multicolumn{5}{|c|}{Model M1b, $\xi$ = 15 \% }\\
\hline
\hline
Sink & $M_{\rm star}$ (M$_{\odot}$) & $M_{\rm disc}$ (M$_{\odot}$) & $R_{\rm disc}$ (au) & $M_{\rm disc}$/$M_{\rm star}$ \\
%\tableline
\hline
1   & 0.117  &   0.038    &   14.914 & 0.324  \\
2   & 0.243  &   0.054    &   14.962 & 0.223  \\
8   & 0.081  &   0.007    &   14.276 & 0.087  \\
\hline
%\tableline
\multicolumn{5}{|c|}{Model M2b, $\xi$ = 15 \% }\\
\hline
\hline
Sink & $M_{\rm star}$ (M$_{\odot}$) & $M_{\rm disc}$ (M$_{\odot}$) & $R_{\rm disc}$ (au) & $M_{\rm disc}$/$M_{\rm star}$ \\
%\tableline
\hline
1   & 0.203  &   0.071    &   19.537 & 0.351  \\
6   & 0.174  &   0.048    &   17.916 & 0.280  \\
7   & 0.204  &   0.010    &   6.879  & 0.050  \\
11  & 0.168  &   0.017    &   9.949  & 0.103  \\

\hline

\multicolumn{5}{|c|}{Model M3b, $\xi$ = 15 \% }\\
\hline
\hline
Sink & $M_{\rm star}$ (M$_{\odot}$) & $M_{\rm disc}$ (M$_{\odot}$) & $R_{\rm disc}$ (au) & $M_{\rm disc}$/$M_{\rm star}$ \\
%\tableline
\hline
2   & 0.129  &   0.118   &   29.849  & 0.920 \\
6   & 0.175  &   0.013   &   29.746  & 0.078 \\
9   & 0.203  &   0.042   &   29.442  & 0.207 \\
11  & 0.089  &   0.015   &   19.828  & 0.168 \\
16  & 0.048  &   0.033   &   19.681  & 0.686 \\
17  & 0.080  &   0.008   &   9.453   & 0.108 \\
24  & 0.022  &   0.004   &   9.829   & 0.188 \\
\hline
%\tableline

\multicolumn{5}{|c|}{Model M4b, $\xi$ = 15 \% }\\
\hline
\hline
Sink & $M_{\rm star}$ (M$_{\odot}$) & $M_{\rm disc}$ (M$_{\odot}$) & $R_{\rm disc}$ (au) & $M_{\rm disc}$/$M_{\rm star}$ \\
%\tableline
\hline
1   & 0.122  &   0.070    &   14.922  & 0.578  \\
6   & 0.198  &   0.005    &   9.789   & 0.029  \\
8   & 0.207  &   0.009    &   9.517   & 0.047  \\
13  & 0.091  &   0.014    &   14.017  & 0.161  \\
17  & 0.054  &   0.003    &   14.920  & 0.064  \\
26  & 0.046  &   0.022    &   19.685  & 0.494  \\
29  & 0.020  &   0.006    &   14.819  & 0.324 \\
%31  & 0.006  &   0.0002   &   4.633   & 0.033 & iso & $-$ & yes \\
%32  & 0.003  &   0.005    &   9.897   & 1.860 & iso & 25$^{\circ}$ & yes\\
\hline
%\tableline

\end{tabular}
%\end{flushleft}
\end{table}

\begin{table} \label{tbl-1}
%\begin{flushleft}
\centering
\caption{Summary of the models in set M5b$-$M8b with the second random seed (corresponding to two different realisations of turbulence, with the same statistical properties). The table is constructed for sink particles (protostars) that form around them a CD structure. Only Class 0 objects are included in the table. The entire table is constructed at the points in time when SFE $\xi$ reaches 15 \% in each model except models M7b and M8b where $\xi$ is 10 \%. The table describes the identity of protostar (Sink), the mass of the protostar ($M_{\rm star}$), the mass of the disc ($M_{\rm disc}$), the radius of the disc ($R_{\rm disc}$), and the disc$-$star mass ratio ($M_{\rm disc}$/$M_{\rm star}$). The mass and the radius are provided in solar mass units and in au, respectively.   }
%\headline{$\xi$ = 2 \%}

%. $\parallel$. $\perp$ symbols for parallel and perpendicular

\begin{tabular}{ccccc}
\multicolumn{5}{|c|}{Model M5b, $\xi$ = 15 \% }\\
\hline
\hline
Sink & $M_{\rm star}$ (M$_{\odot}$) & $M_{\rm disc}$ (M$_{\odot}$) & $R_{\rm disc}$ (au) & $M_{\rm disc}$/$M_{\rm star}$ \\
%\tableline
\hline
1   & 0.224  &   0.023    &   19.971 & 0.003  \\
2   & 0.407  &   0.003    &   18.416 & 0.002  \\
3   & 0.085  &   0.029    &   39.387  & 0.064  \\
\hline

\multicolumn{5}{|c|}{Model M6b, $\xi$ = 15 \% }\\
\hline
\hline
Sink & $M_{\rm star}$ (M$_{\odot}$) & $M_{\rm disc}$ (M$_{\odot}$) & $R_{\rm disc}$ (au) & $M_{\rm disc}$/$M_{\rm star}$ \\
%\tableline
\hline
%2   & 0.256  &   0.001    &   9.828 & 0.003  \\
3   & 0.063  &   0.005    &   14.618  & 0.087  \\
4   & 0.146  &   0.019    &   12.449  & 0.134  \\

\hline
\multicolumn{5}{|c|}{Model M7b, $\xi$ = 10 \% }\\
\hline
\hline
Sink & $M_{\rm star}$ (M$_{\odot}$) & $M_{\rm disc}$ (M$_{\odot}$) & $R_{\rm disc}$ (au) & $M_{\rm disc}$/$M_{\rm star}$ \\
%\tableline
\hline
1   & 0.284  &   0.016    &   12.646 & 0.056  \\
2   & 0.098  &   0.013    &   18.689 & 0.139  \\
6   & 0.028  &   0.002    &   17.962  & 0.074 \\
\hline

\multicolumn{5}{|c|}{Model M8b, $\xi$ = 10 \% }\\
\hline
\hline
Sink & $M_{\rm star}$ (M$_{\odot}$) & $M_{\rm disc}$ (M$_{\odot}$) & $R_{\rm disc}$ (au) & $M_{\rm disc}$/$M_{\rm star}$ \\
%\tableline
\hline
1   & 0.398  &   0.002    &   19.548 & 0.004  \\
4   & 0.093  &   0.006    &   18.441 & 0.064  \\
7   & 0.050  &   0.002    &   34.747  & 0.037  \\

\hline
\end{tabular}
%\end{flushleft}
\end{table}

\begin{figure}
	 %To include a figure from a file named example.*
	% Allowable file formats are eps or ps if compiling using latex
	% or pdf, png, jpg if compiling using pdflatex
%	\includegraphics[width=\columnwidth]{myradiusmass1.png}
%	\includegraphics[width=\columnwidth]{UPDATEDspecific_AM_seed2_M1b_M8b.png}
 \includegraphics[width=\columnwidth]{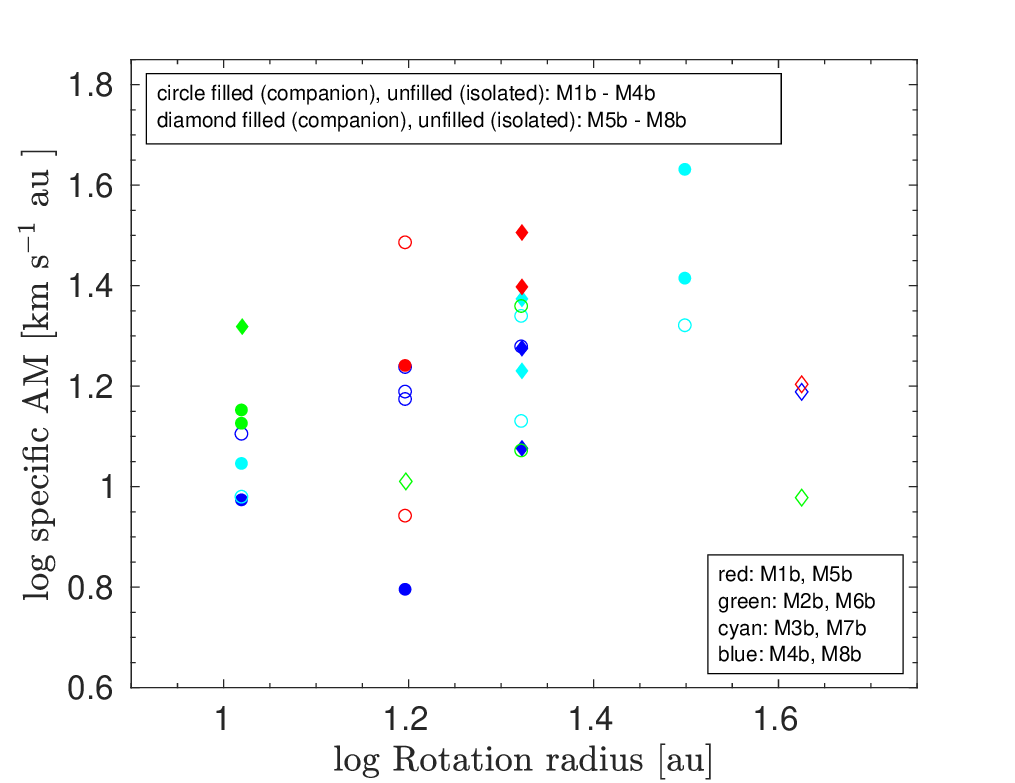}
 
	\caption{Radial distribution of specific AM for models M1b$-$M8b at the end of our computation when star formation efficiency (SFE) in each model reaches $\xi$ = 15 \%, except models M7b and M8b where simulations due to high computational cost are terminated at $\xi$ = 10 \%.  Units of specific AM and disc radius are provided in km s$^{-1}$ au and au, respectively. Both quantities are presented in the logarithmic scales. Filled and unfilled circles indicate the discs which are part of the binary systems and the isolated systems, respectively in model set M1b$-$M4b. Similarly, filled and unfilled diamonds indicate the discs which are part of the binary systems and the isolated systems, respectively in model set M5b$-$M8b. Colour red, green, cyan, and blue represent models M1b and M5b, M2b and M6b, M3b and M7b, and M4b, and M8b, respectively. Colour in the online edition.}
	\label{fig:figur6}
\end{figure}

\begin{figure}
	 %To include a figure from a file named example.*
	% Allowable file formats are eps or ps if compiling using latex
	% or pdf, png, jpg if compiling using pdflatex

%	\includegraphics[angle=0,scale=0.5675]{REVISED_sigmaprofile_seed2_M1b_M4b.eps}
%	\includegraphics[angle=0,scale=0.5675]{REVISED_sigmaprofile_seed2_M5b_M8b.eps}
 \includegraphics[width=\columnwidth]
  {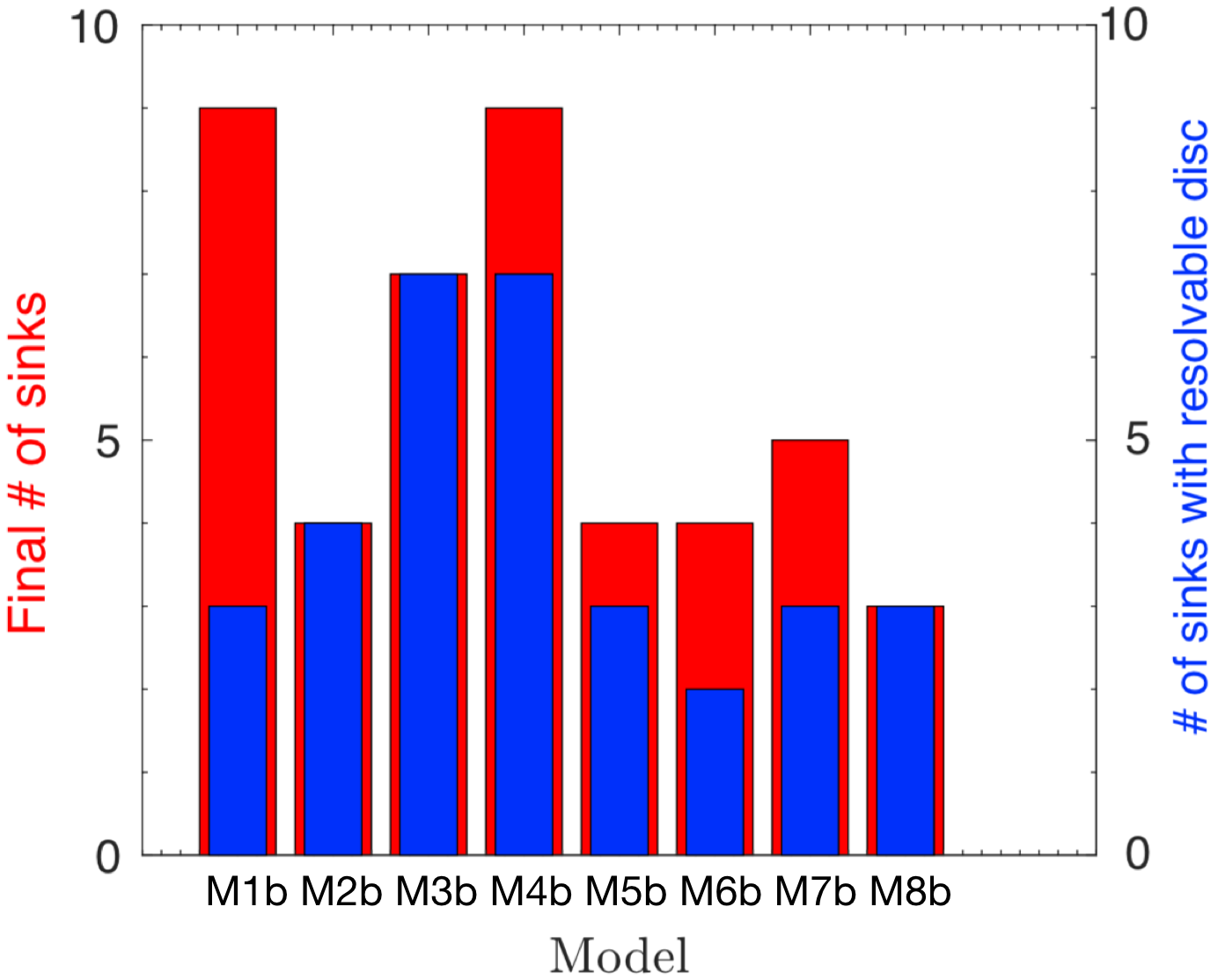}
	\caption{The relative disc abundance i.e. the number of sinks with resolved discs (blue histogram) compared to the total number of surviving sinks (red histogram) when SFE reaches $\xi$ = 15 \% , except models M7b and M8b where simulations due to high computational cost are terminated at $\xi$ = 10 \%. Colour in the online edition.}
	\label{fig:figur6}
\end{figure}

\begin{figure*}
	 %To include a figure from a file named example.*
	% Allowable file formats are eps or ps if compiling using latex
	% or pdf, png, jpg if compiling using pdflatex

%	\includegraphics[angle=0,scale=0.5675]{REVISED_sigmaprofile_seed2_M1b_M4b.eps}
%	\includegraphics[angle=0,scale=0.5675]{REVISED_sigmaprofile_seed2_M5b_M8b.eps}
 \includegraphics[angle=0,scale=0.525]
  {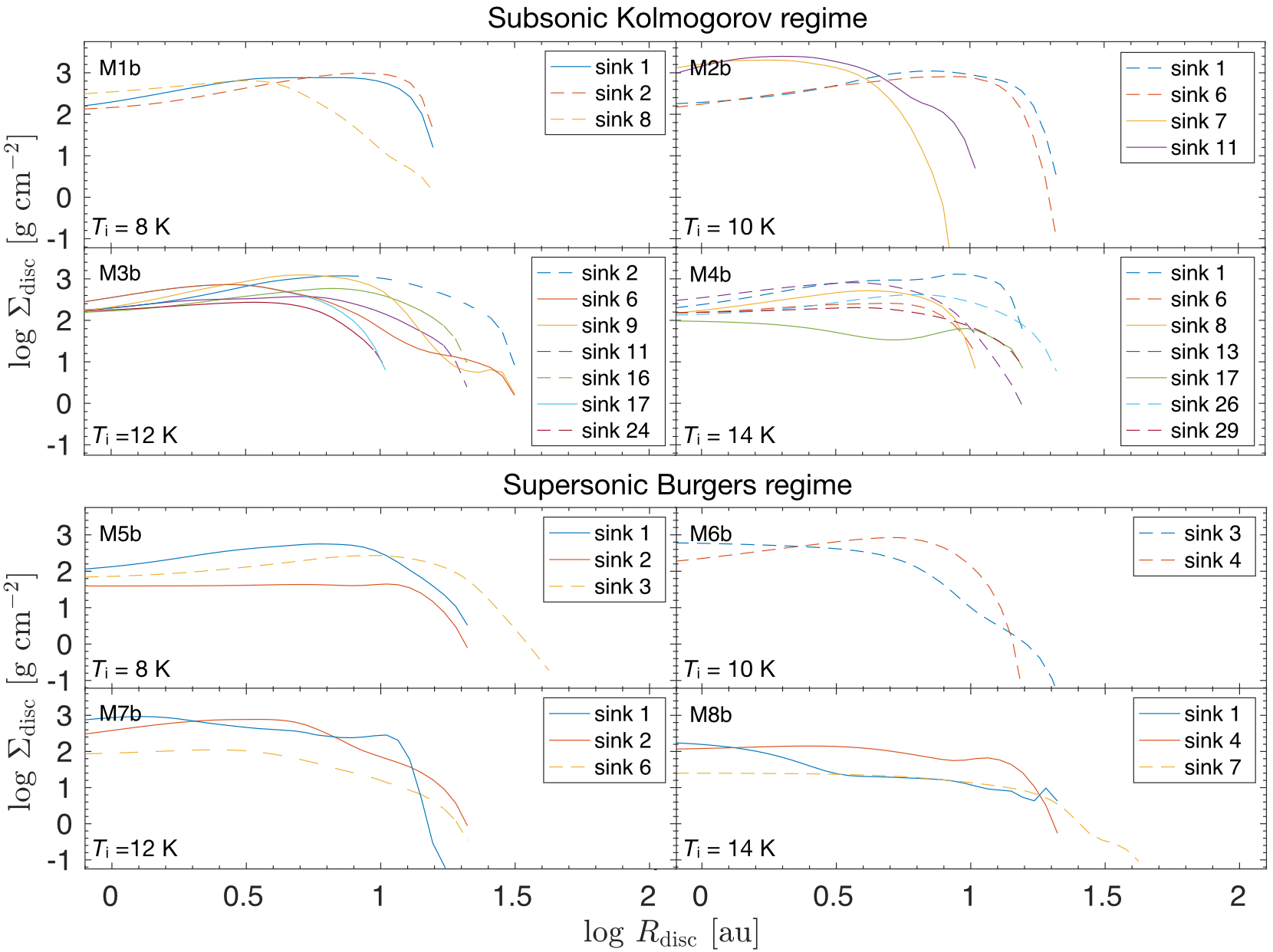}
	\caption{Radial distribution of surface density of the discs associated with sink (protostar). The top four panels are for models M1b$-$M4b and the bottom four panels are for models M5b$-$M8b. Each radial surface density curve in all respective eight panels is shown with a distinct colour where the solid line indicates disc radial profile of protostar that is a companion of a binary system and the dashed line indicates isolated protostar. The radius and the surface density are in the logarithmic scales and are given in units of au and g cm$^{-2}$, respectively. Colour in the online edition.}
	\label{fig:figur6}
\end{figure*}

\begin{figure*}
	 %To include a figure from a file named example.*
	% Allowable file formats are eps or ps if compiling using latex
	% or pdf, png, jpg if compiling using pdflatex
%	\includegraphics[angle=0,scale=0.575]{UPDATEDVradialprofile_seed2_M1b_M4b.eps}
%	\includegraphics[angle=0,scale=0.575]
% {UPDATEDVradialprofile_seed2_M5b_M8b.eps}

\includegraphics[angle=0,scale=0.525]
  {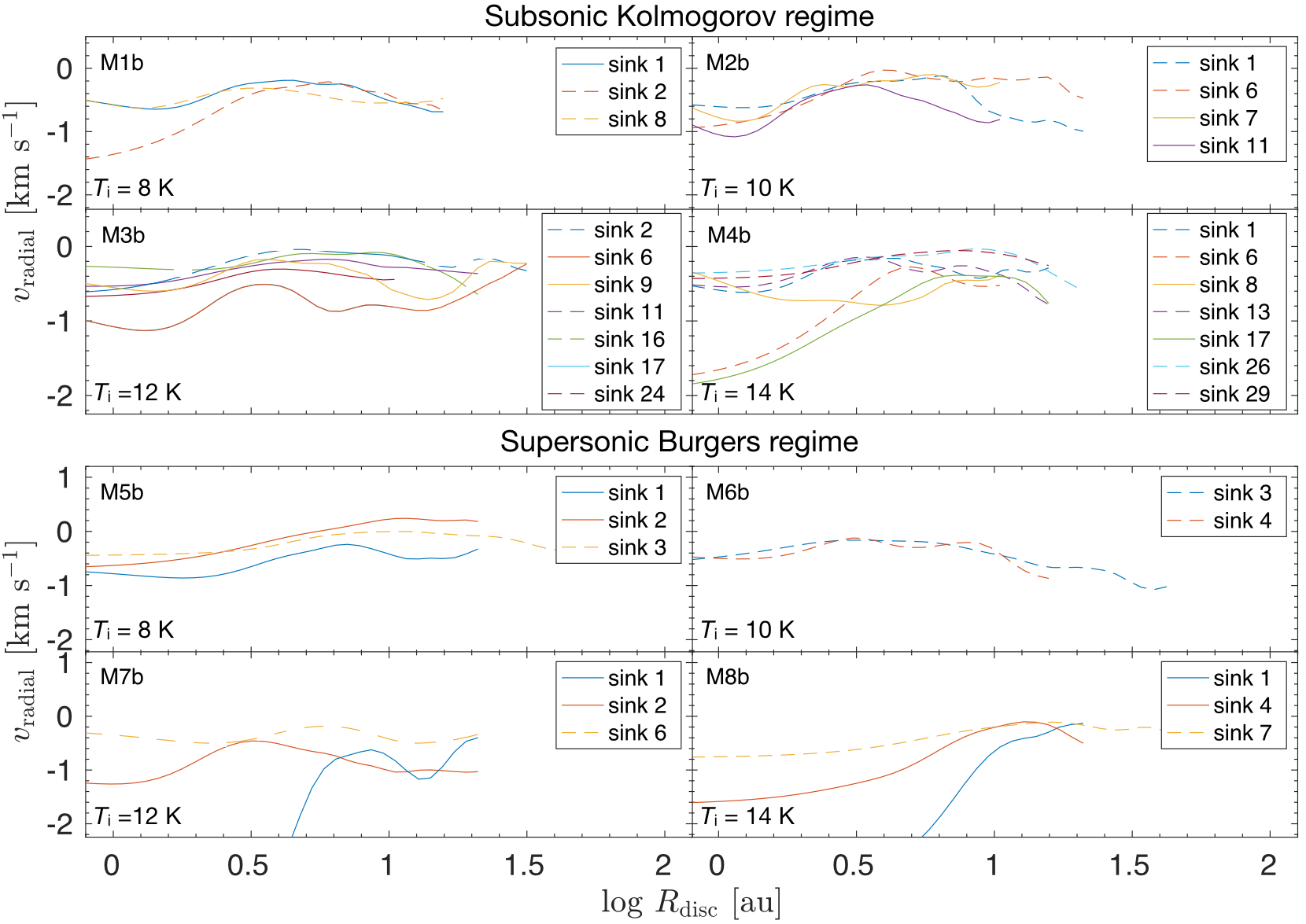}
 
	\caption{Radial velocity distribution of the  discs associated with sink (protostar). The top four panels are for models M1b$-$M4b and the bottom four panels are for models M5b$-$M8b. Each radial velocity curve in all respective eight panels is shown with a distinct colour where the solid line indicates disc radial profile of protostar that is a companion of a binary system and the dashed line indicates isolated protostar. The radius is in the logarithmic scales. The radius and the radial velocity are given in units of au and km s$^{-1}$, respectively. Colour in the online edition.}
	\label{fig:figur6}
\end{figure*}

\begin{figure*}
	 %To include a figure from a file named example.*
	% Allowable file formats are eps or ps if compiling using latex
	% or pdf, png, jpg if compiling using pdflatex
%	\includegraphics[angle=0,scale=0.575]{UPDATEDVaziprofile_seed2_M1b_M4b.eps}
%	\includegraphics[angle=0,scale=0.575]{UPDATEDVaziprofile_seed2_M5b_M8b.eps}

 \includegraphics[angle=0,scale=0.525]
 {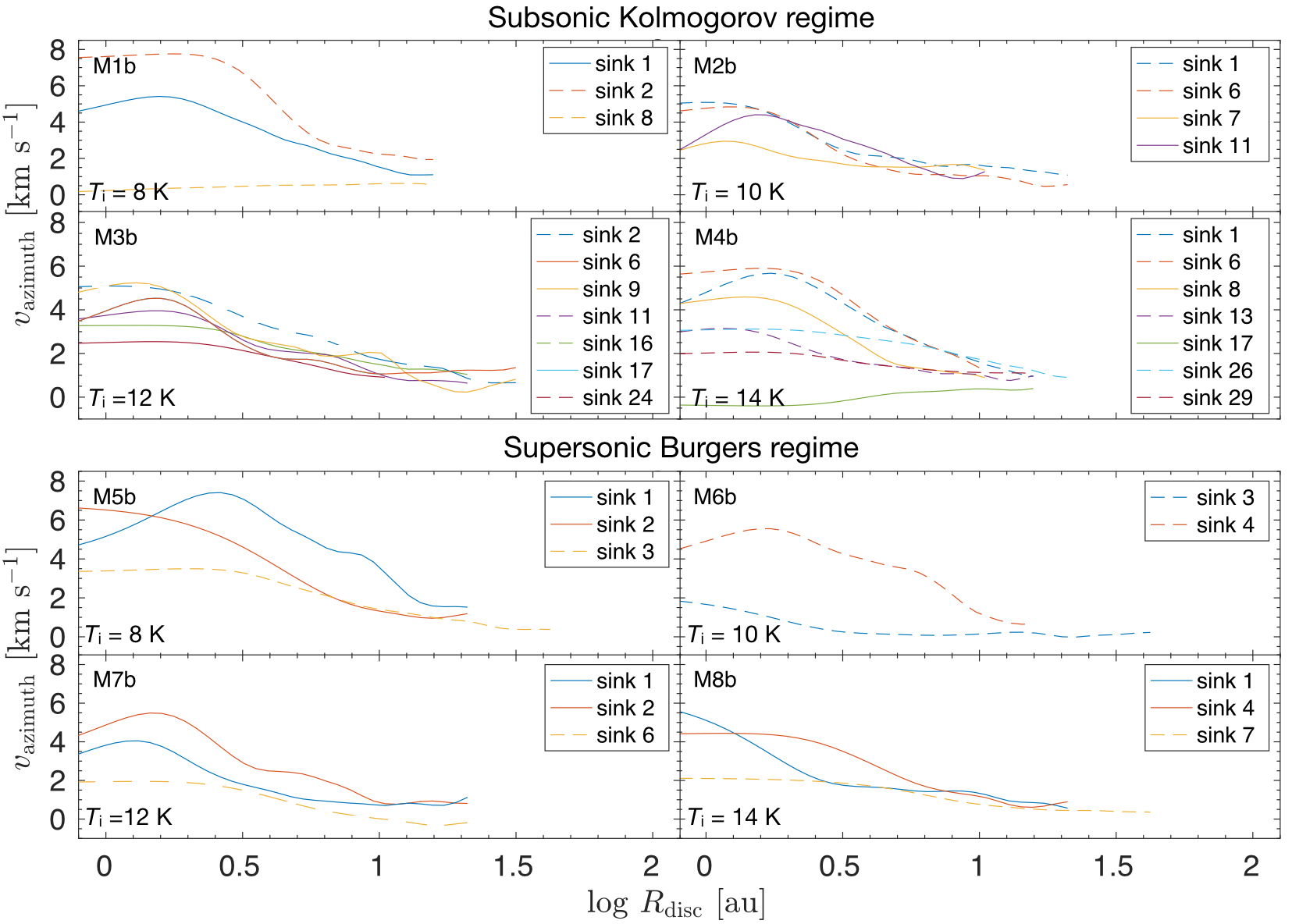}
	\caption{Azimuthal velocity distribution of the  discs associated with sink (protostar). The top four panels are for models M1b$-$M4b and the bottom four panels are for models M5b$-$M8b. Each azimuthal velocity curve in all respective eight panels is shown with a distinct colour where the solid line indicates disc radial profile of protostar that is a companion of a binary system and the dashed line indicates isolated protostar. The radius is in the logarithmic scales. The radius and the azimuthal velocity are given in units of au and km s$^{-1}$, respectively. Colour in the online edition.}
	\label{fig:figur6}
\end{figure*}

\begin{figure*}
	 %To include a figure from a file named example.*
	% Allowable file formats are eps or ps if compiling using latex
	% or pdf, png, jpg if compiling using pdflatex
%	\includegraphics[angle=0,scale=0.5675]{UPDATEDMach_seed2_M1b_M4b.eps}
%	\includegraphics[angle=0,scale=0.5675]{UPDATEDMach_seed2_M5b_M8b.eps}

 \includegraphics[angle=0,scale=0.525]{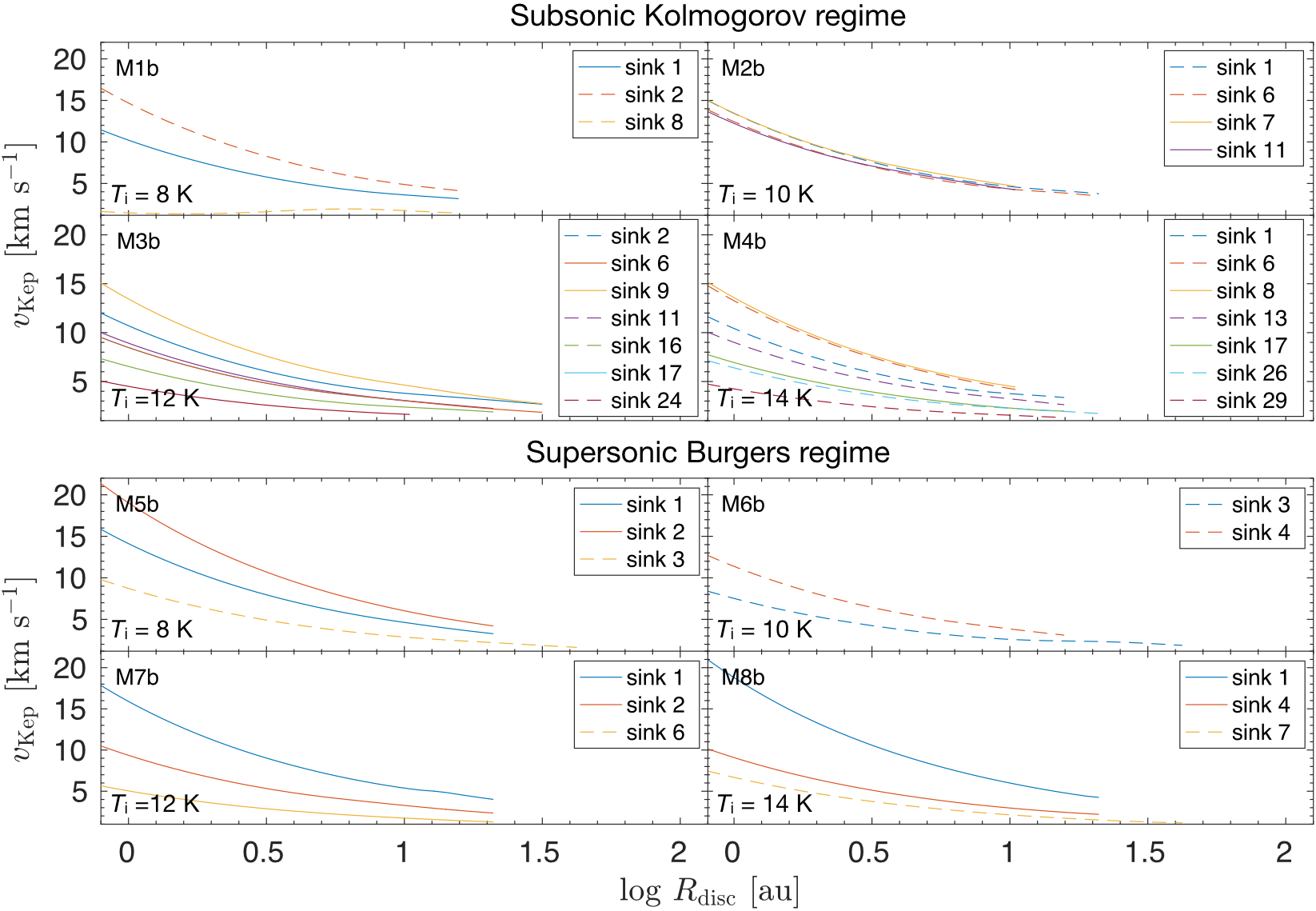}
	\caption{Keplerian velocity distribution of the discs associated with sink (protostar). The top four panels are for models M1b$-$M4b and the bottom four panels are for models M5b$-$M8b. Each Keplerian velocity curve in all respective eight panels is shown with a distinct colour where the solid line indicates disc radial profile of protostar that is a companion of a binary system and the dashed line indicates isolated protostar. The radius is in the logarithmic scales. The radius and the Keplerian velocity are given in units of au and km s$^{-1}$, respectively. Colour in the online edition.}
	\label{fig:figur6}
\end{figure*}

\begin{figure*}
	 %To include a figure from a file named example.*
	% Allowable file formats are eps or ps if compiling using latex
	% or pdf, png, jpg if compiling using pdflatex
%	\includegraphics[angle=0,scale=0.5675]{UPDATEDqprofile_seed2_M1b_M4b.eps}
%	\includegraphics[angle=0,scale=0.5675]{UPDATEDqprofile_seed2_M5b_M8b.eps}

 \includegraphics[angle=0,scale=0.525]{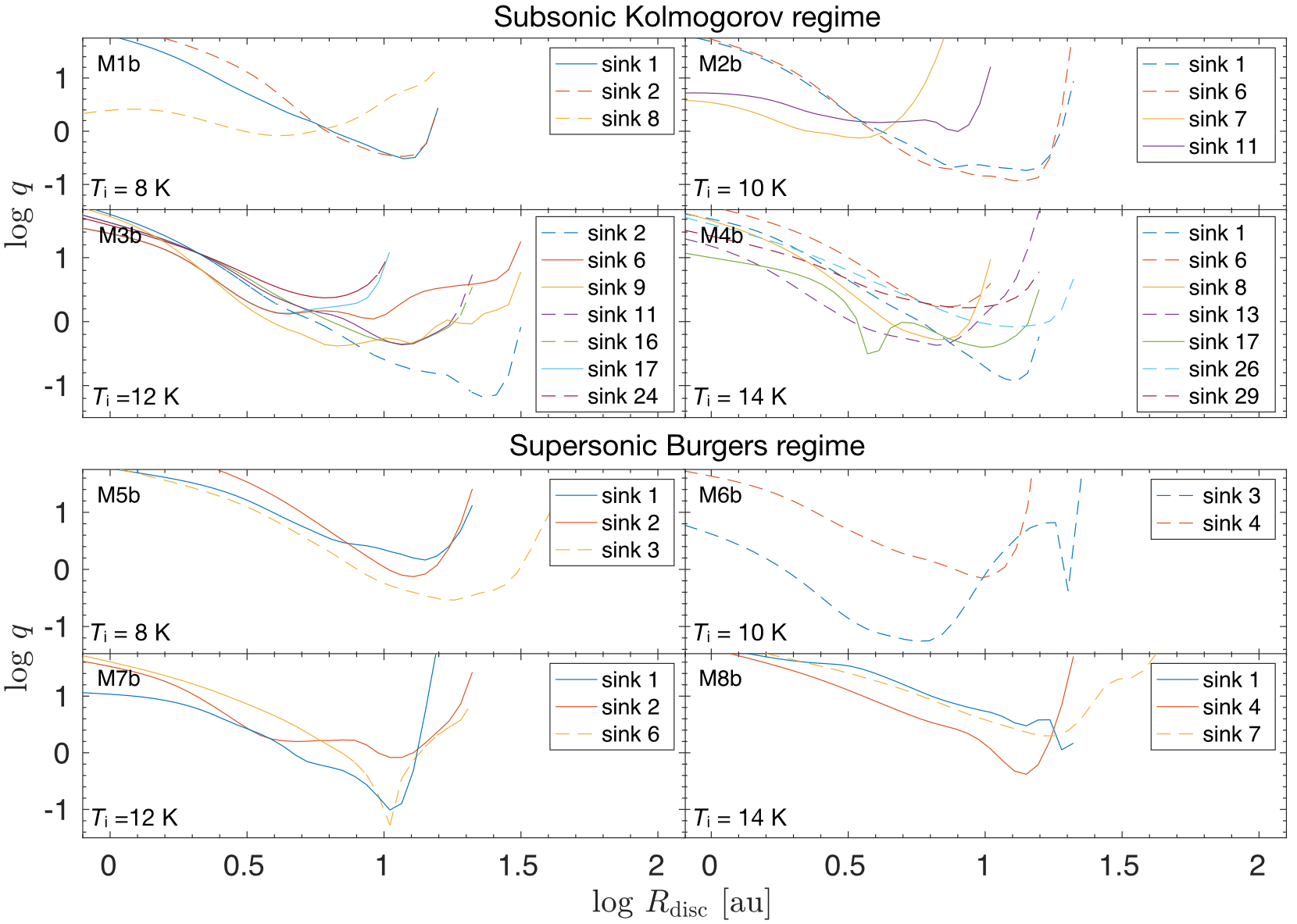}
	\caption{$q$ $-$ parameter profile of the discs associated with sink (protostar). The top four panels are for models M1b$-$M4b and the bottom four panels are for models M5b$-$M8b. Each  curve in all respective eight panels is shown with a distinct colour where the solid line indicates disc radial profile of protostar that is a companion of a binary system and the dashed line indicates isolated protostar. The radius is in the logarithmic scale and is given in units of au. Colour in the online edition.}
	\label{fig:figur6}
\end{figure*}
\vspace{-0.5cm}
\section{Results and discussion}
We now present our simulation results with a focus on the CDs produced in our models and discuss them in detail. For the visualisation of our simulation results, we use the visualisation tool SPLASH developed by \citet{price2007splash}.
\vspace{-0.5cm}
\subsection{Turbulent fragmentation and the formation of circumstellar discs }
Most prestellar cores are observed to have subsonic velocity dispersion \citep{myers1983dense, jijina1999dense, bergin2007cold} with only a small fraction reaching transsonic values (for Serpens, see  \citep{williams1999contracting}. Therefore, we numerically investigate turbulent fragmentation in prestellar gas cores and its effect on the formation of circumstellar discs around protostars.
The evolution from the initial collapse to the formation of a protostar involves intricate physics, including the role of temperature changes, dissociation of molecules like $H_{\rm 2}$, accretion of material onto the protostar, and the formation of a CD. The differentiation in AM within the collapsing material is crucial. It contributes to the formation of the CD around the protostar once the rotational velocity becomes equal to the Keplerian velocity. The presence of material with higher AM tends to settle into a rotating disc structure, which gradually accretes onto the protostar. The turbulent nature of the collapsing gas can lead to a variety of orientations for these CDs. The variations in the orientation of these discs might impact the eventual arrangement and characteristics of the planets that form within these systems.
In addition to the turbulence, an important factor that may influence the process of fragmentation and the subsequent formation of circumstellar discs is the ratio of rotational energy to gravitational energy ($\beta$). Our choice of $\beta$ in the present work is closely related to the observed lower range of $10^{-4}$  < $\beta$ < 0.07 in the sample studies by \citet{caselli2002dense}. Moreover, the present study is closely related to the hydrodynamic cases explored by \citet{wurster2020non} in terms of the initial mass and size of the prestellar gas cores. However, in their study, they introduced sink particles in the collapsing gas at a density which is threshold three orders of magnitude greater than ours. 
They have also investigated both subsonic and transsonic turbulence and the subsequent effect on circumstellar discs. The higher density threshold set for the formation of sink particles in the study of \citep{wurster2020non} has resulted in a lower number of protostars and hence a subsequent lower number of circumstellar discs as compared to the outcome of our simulations. On the other hand, we consider subsonic and supersonic turbulent motions inside the prestellar gas core where the former is modelled with Kolmogorov-type and the latter is modelled with Burgers-type turbulent velocity spectra. More efficient turbulent fragmentation resulted in the case of Kolmogorov-type turbulent velocity spectrum than in the case of Burgers-type turbulent velocity spectrum. This has translated into more extended and comparably more massive discs in the subsonic, Kolmogorov regime than in the supersonic, Burgers regime (see Figure 2).

Because of the greater number of computationally resolvable CDs produced in our models that are based on the second seed i.e. M1b$-$M8b, we  particularly focus on the analysis of this model set and discuss its outcome in detail. Our model results for the first seed M1a$-$M8a are briefly presented in Appendix B for readers to understand this biased selection.

\citet{wurster2019there} have found that protostellar discs with radii around 20 au form rather frequently. Their findings indicate that disc formation is regulated rather by gas turbulence than magnetic fields, suggesting an interesting interplay of forces during the collapse of the prestellar gas core. The lack of evidence for the "magnetic braking catastrophe" aligns with the idea that magnetic fields might not significantly impede or hinder the formation of these protostellar discs. Their result can be explained by two factors: 1) the turbulent gas motions result in twisting of the magnetic fiel lines, which in turn leads to reconnection and a drastical reduction of the efficiency of magnetic braking as a result of the increasing complecity of the topology of the magnetic field.  2) The presence of non-ideal MHD effects like Ohmic dissipation, ambipolar diffusion and the Hall effect can weaken or decouple magnetic field from the gas, again reducing the magnetic braking effect.  The strong magnetic braking effects on disc formation reported by \citet{price2007impact} are therefore not necessarily relevant for disc formation in turbulent molecular cloud cores because their simulations used a simple vertical magnetic field in their initial conditions and neglected turbulent motions within the gas and the impact of the non-ideal MHD terms in the induction equation. 
The prominence of gas turbulence in facilitating the formation of discs implies that other factors, such as the initial conditions of the collapsing gas and the specific properties of the turbulent flow (as explored in this work) will play critical roles in shaping the outcome of the collapse and subsequent disc formation.
In Figure 2, we show histograms of the disc masses (top panel) and disc radii (bottom panel) of Kolmogorov-type turbulence (blue) and Burger-type turbulence (red) in model sets M1b$-$M4b and M5b$-$M8b, respectively. We discuss these two types of models of turbulence without taking into account the initial thermal states of the collapsing molecular gas cores. We do not classify the isolated and binary configurations while showing these distributions. In the top panel, we show that our models based on Kolmogorov-type turbulence yield  more massive discs than Burgers-type models. However, on the lower end of the disc mass-spectrum, less massive discs are produced in the former type of models than in the latter. This can be explained by taking into account the identical SFE reached in both types of models, and with a fewer number of CDs produced in Burgers type of models. In general, we find in the star-forming gas a wider range of CD masses in the case of Kolmogorov turbulence but not for Burgers type of turbulence. Gravitational torques may dominate in the early stages when the disc mass is greater than or comparable to the stellar mass.  In contrast, viscous effects may induce a more gentle but steady accretion that will be the primary mover when the disc mass is less than the stellar mass \citep{basu1998constraints}. In this context, we believe that most of the CDs in our models are under the influence of viscous effects rather than gravitational torques (see Tables 2 and 3). In Figure 2 (bottom panel), we show the number of CDs as a function of disc radii. In the case of molecular core with Kolmogorov-type turbulence, discs with smaller radii are formed than in molecular core with Burgers-type turbulence. Also, a significant number of  discs are formed with a radius of $\sim$ 15 au in both cases of Kolmogorov and Burgers types of turbulence in star-forming gas. However, a greater number of extended discs with radii $>$ 15 au is yielded by the former type of turbulent models than the latter.

It is worth highlighting that the small size discs around YSOs may not necessarily be a result of magnetic braking and that they could be produced hydrodynamically. For instance, discs may be born small due to turbulent fragmentation. Comparing the decay time of turbulence to the free-fall time of the gas can provide insights of whether disc sizes are set by initial turbulent fragmentation. If turbulence decays much quicker than the gas collapse, it suggests disc sizes are not determined by turbulent fragmentation. Moreover, discs may be born at various sizes, but close interactions within the core tidally truncate them to smaller sizes. \citet{winter2018protoplanetary} found that the truncated radius is related to the distance of closest approach. Any encounters within certain distances could truncate discs to specific sizes. Also, disc size evolution may be limited by the stochastic nature of angular momentum evolution in the protostellar environment. Studies like \citet{kuznetsova2019origins} observe stochastic angular momentum evolution at the core scale, which could influence disc size evolution. Each of these pathways offers potential explanations for the small size of CDs in protostars.

On the basis of calculations performed until $10^{5}$ year after the protostar formation, \citep{goodwin2004simulating} showed that fragmentation frequently occurs even in a weakly turbulent environment. By taking the initial structure of the simulated core similar to the observed properties of the A-MM8 core in Ophiuchus, \citep{walch2012influence} have demonstrated in their numerical study that the dominant mode of fragmentation involves the formation and break-up of filaments and although small protostellar discs (with radius less than or equal to 20 au) form routinely, more extended discs are rare. Furthermore, in turbulent, low-mass cores of the type simulated by \citet{walch2012influence}, the formation of large fragmenting protostellar discs is suppressed by early fragmentation in the filaments. The formation and evolution process of circumstellar discs in turbulent cloud cores differs substantially from that in rigidly rotating cloud cores \citep{tsukamoto2013formation}. In the turbulent fragmentation scenario, turbulent perturbations developing inside the prestellar gas core or within the filaments collapse distinctively to form multiple protostars. If the turbulence is initially sufficiently strong, the remaining filaments twist around the protostars and directly become rotation-supported discs \citep{tsukamoto2013formation}. Upon formation, the disc orientation is generally misaligned with the AM of its host cloud core and it dynamically varies during the main accretion phase, even though the turbulence is weak. This is due to the AM of the entire cloud core, which is mainly determined by the large scale velocity field whose wavelength is comparable to the cloud scale, whereas the AM of the disc is determined by the local velocity field where the protostar forms. These two velocity fields do not generally correlate with each other. 

%\subsubsection{Disc mass and radius distributions}
\citet{greaves2011all} have provided estimates for the disc radii and masses of Class 0 protostars which show a range over 10 $-$ 200 au and 7 $-$ 660 $M_{\rm Jup}$, respectively. Later on, \citet{yen2015observations} 
suggested for Class 0/I that disc radii range from < 5 au to > 500 au with estimated protostellar masses from < 0.1 M$_{\odot}$ to > 1 M$_{\odot}$. Also, more recently several Class 0 protostars seem to have very small discs with radius as small as 3 au \citep{bjerkeli2019kinematics}.

When comparing two or more data distributions, it often requires statistical tests to be conclusive about the significant differences. However, the selection of these tests usually depends highly on the intrinsic properties of those distributions \citep[e.g.,][]{Skovlund2001}. Due to the amount of simulation data available to us for both  turbulence cases, we perform a Student's $t$-test \citep{Student1908} to determine if there is a significant difference between the means of the two distributions and a Mann-Whitney $U$-test \citep{Mann1947} to compare whether or not the samples are from the same underlying population.

The $t$-test computes the $t$ statistic by taking the difference between the means of the two groups and dividing it by the standard error of the difference, which is influenced by the sample sizes and variances within each group. This $t$ statistic follows a $t$-distribution under the null hypothesis of no difference between the groups. Conversely, the Mann-Whitney $U$-test calculates the $U$ statistic by comparing the ranks of the observations between the two groups. It sums the ranks of the observations from one group and compares it to the sum of ranks from the other group. It tests whether the distributions of the two groups differ significantly from each other, regardless of the shape of the distributions.

The results of the $t$-test and the $U$-test between the mass and radius distributions can be seen on Figure 3 in top and bottom panels, respectively. The boxes represent the interquartile range (IQR) measured from the 25 to the 75 percentile, while the whiskers represent the maximum an minimum values expected around the IQR. Any sample outside the whiskers is taken as an outlier. The $p$-value obtained from the $t$-test for the mass and radius distributions allows us to reject the hypothesis of equality of means between the two types of turbulence with a significance level of approximately 5\% (or 95\% confidence). Meanwhile, the p-value obtained from the $U$-test for the mass and radius distributions allows us to reject the hypothesis of equality between the two distributions with a significance level of around 9\% (or 91\% confidence). Therefore, if we relax our margin of error to 9\%, we can conclude that the distributions for the masses and the radii of protoplanetary discs will change depending on the type of turbulence we choose.

In Figure 4, we show disc radii as a function of disc masses for circumstellar disc structures formed in our models M1b$-$M8b. We find for Class 0 objects the disc radii in the range of $\sim$ 7 $-$ 32 au and $\sim$ 9 $-$ 35 au in Kolmogorov-type of turbulence (i.e. M1b$-$M4b) and Burgers type of turbulence (i.e. M5b$-$M8b), respectively. The mass range for Class 0 objects is obtained as 30.37 $M_{\rm Jup}$ $-$ 0.92 M$_{\odot}$, and 2.09 $M_{\rm Jup}$ $-$ 0.13 M$_{\odot}$ in Kolmogorov type of turbulence (i.e. M1b$-$M4b) and in Burgers type of turbulence (i.e. M5b$-$M8b), respectively. We do not find any correlation between the disc radius and its mass over the explored range of initial thermal states and the type of turbulence of the collapsing gravoturbulent gas cores.

It has been suggested that close companion objects in the close proximity can affect the size of the CD as a nearby companion(s) can make the disc small by the effect know as tidal truncation \citep{larwood1996tidally, rodriguez2005iras, mcclure2008sub}. In the case of isolated discs, their compactness could be a result of the youth of the discs and not of tidal truncation. As discussed before, small disc sizes do not require the presence of a magnetic field, but disc formation and the resulting disc size also depends on the gas turbulence \citep{wurster2019there}. We also find CDs with small radii and most of these systems are found located very close to each other (see especially Figure 5 for models M1b, M3b, and M4b). Protostars on wide orbits can significantly impact the structure and size of the discs. Fly-by interactions or close encounters with these objects can lead to tidal forces that distort the shape of the discs or strip away material from their outer regions (see also \citep{winter2018protoplanetary}). Tidal stripping occurs when the gravitational forces from the passing stars pull material away from the discs, diminishing their size. In extreme cases, particularly strong interactions might completely disrupt the discs and can scatter their components into different trajectories. Nonetheless, there are a few disc structures which are more isolated but yet show small disc radii (see Figure 6 for models M5b, M6b, and M7b).

\subsection{Disc-to-star mass ratio ($M_{\rm disc}$/$M_{\rm star}$)}
Class 0 protostars are generally low bolometric temperature sources and hence are subject to significant amounts of extinction. Thus, it has been difficult to infer whether these sources have rotationally supported discs, or whether magnetic braking at
the early stages of these systems prevents a disc structure to form (see e.g. \citep{allen2003collapse, mellon2008magnetic, li2013does, wurster2016can, wurster2019there}. Moreover, rotationally supported discs
have been observed around some Class 0 objects (see \citealt{tobin2013modeling, codella2014alma, gerin2017evidence, sakai2019warped, kawasaki2021growth}). For Class 0 and I and even for brown dwarfs, the observed mass ratios between the disc and central protostar are typically $\sim$ 1 \% (see e.g. \citealt{andrews2005circumstellar, scholz2006exploring}), which indicates an efficient mechanism to transport AM outward and mass inward already at an early stage during the life of the disc and even while it is still forming \citep{vorobyov2007self}. Our simulation results based on the second random seed(i.e. M1b$-$M8b) yield a  number of well resolved CDs. We find a systematic trend which suggests that for the resolvable CDs which form under a solenoidal mode of subsonic turbulence, the ratio $M_{\rm disc}$/$M_{\rm star}$ is larger compared to the situation where the discs are formed due to a compressional turbulence mode in supersonic turbulent cores. The ratio $M_{\rm disc}$/$M_{\rm star}$ is also largely determined by the AM of the parent gas core \citep{visser2010sub}. For the gravity-driven turbulence in a contracting core where the driving scale is relatively small as compared to size of the molecular clouds, \citet{xu2020turbulence} have demonstrated that the turbulent motions inside the dense part of the core provide pressure support and hence can slow down the gravitational infall and the subsequent mass accretion. \citet{machida2010formation} have demonstrated the possibility that mass of the CD is comparable to or larger than the
protostar ($M_{\rm disc}$/$M_{\rm star}$ $\gtrsim$ 1) during the Class 0 or Class I stages in their long-term simulations of the main accretion phase of star formation. In our model results M1a$-$M8a and M1b$-$M8b, resolvable discs do not exhibit a disc-star mass ratio greater than unity (see Tables 2 and 3). 

\vspace{-0.5cm}
\subsection{Material infall onto the disc edge}
The origin of core rotation and its time evolution are not well understood. \citet{misugi2019origin} have demonstrated that the observed AM distribution of cores as a function of core mass can be understood by the fragmentation of filaments having velocity fluctuations close to the sonic speed and anisotropic, log-normal 1D Kolmogorov power spectra. Comparing their results with observational findings \citep{goodman1993dense,caselli2002dense,tatematsu2016angular} suggests that for a prestellar core mass of 5 M$_{\odot}$ the specific AM is of the order of $10^{20}$ cm$^{2}$ s$^{-1}$. We also performed simulations of a gravoturbulent core collapse for a core mass of 5 M$_{\odot}$. Our simulations show the formation of filamentary structures that host star-formation and the subsequent disc formation around protostars. When the SFE reaches 15 \% during collapse, the specific AM of the circumstellar disc structures, which has its origin in the AM of the parent gas cores, exhibits an average specific AM of the order of $10^{16}$ cm$^{2}$ s$^{-1}$ (we adopt their system of units while quoting this numerical value).
This can be seen in Figure 7 where we show the specific AM as a function of the radius with respect to the disc and the central protostar. Also, \citet{gaudel2020angular} derived the radial distribution of the local specific AM for various sources ranging from molecular clouds to sun-like stars. Their results suggest that the specific AM observed at their explored > 1600 au scales may not be directly related to rotational motions of the envelopes but could have its origin in other mechanisms, such as core-forming motions (i.e. infall and turbulence). Understanding the dynamics of AM redistribution during this stage is crucial for comprehending the formation and evolution of protostellar systems. The specific AM of Class 0 discs derived in their work suggests a value of the order of $10^{-4}$ km s$^{-1}$ pc (see Figure 17, left panel in their paper). Adopting their system of units, the class 0 discs in our simulations have a specific AM of the order of $10^{-6}$ km s$^{-1}$ pc.   

The specific angular momentum of the discs is computed as follows: 

\begin{equation} \label{}
l = v_{\rm \phi} \, r_{\rm disc\,edge},
\end{equation}
where $l$, $v_{\rm \phi}$ and ${r_{\rm disc\,edge}}$ represent the specific AM, the azimuthal velocity at the edge of the disc, and the outer edge of the disc-radius, respectively (see e.g. \citealt{lee2017formation, shariff2022protostellar}).

In this definition of the specific AM of the infalling gas onto the disc we include all protostars and their potential discs in our sample of second seed-based models i.e. M1b$-$M8b.  This gives us an insight into the distribution of the specific AM in the dense regions of the collapsing gas cores. Most of the CDs in our simulations which are associated with isolated and binary companions are formed with a radius of $\sim$ 16 au $-$ 21 au. For systems of discs with their radii in this range, the gas infall from the envelope onto the edge of the discs exhibits extreme variations in specific AM. We believe that these variations are primarily due to the presence of velocity flows of various magnitudes in the collapsing gas core which are caused by the turbulence in the gas (as pointed out in the previous section). Also, more importantly, the disc structure around binary companions (of the same binary system) shows significantly higher specific AM in the infalling material from the surroundings. This is consistent with previous studies showing that in addition to the more massive and extended circumprimary disc, the circumsecondary disc can only be massive and hence extended enough if the infalling material onto the disc carries higher specific AM \citep{bate1997accretion, bonnell1997accretion, riaz2021turbulence}. This is visible In Figure 7, especially for Burgers-type turbulence models in which both components of the binary systems exhibit relatively more massive and extended discs in the presence of the gas infall with high specific AM. We also find for Burgers-type turbulence models that the isolated disc structures show the most extended radii with sufficiently high specific AM gas infall onto the discs. The turbulence promotes filamentary-type of infall from the envelope to the disc. This continuous or episodic process of material infall gives rise to the formation of a disc around the protostar with an inclination which is mainly controlled by the direction of the infall itself. Also, such an infall of material may result in the disc becoming susceptible to gravitational
instability over time \citep{kuznetsova2019origins}. 
%Also, only the isolated discs in the Kolmogorov-type turbulence model are with radii $\sim$ 20 au. While in the Burgers-type turbulence model, both isolated and binary protostars develop CDs with a radius $\sim$ 20 au.

As suggested by \citet{machida2011effect} the difference in the disc radii is due to the difference in the initial distribution of the AM.  Also, the AM of the infalling gas has a strong impact on disc evolution \citep{vorobyov2015effect}. Small discs with a radius of less than 100 au are expected to be produced by the magnetized turbulent model. \citep{matsumoto2017circumstellar}. Also, \citet{lee2018alma} has shown that a Keplerian disc can form around protostars with a radius as small as $\sim$ 10 au. We have demonstrated in our hydrodynamical simulation models that an unmagnetised picture of the prestellar gas core under gravoturbulent gas collapse can still produce CDs with radii in the range of $\sim$ 7 au to $\sim$ 40 au, which is consistent with previous studies \citep{yen2013unveiling,yen2015no,lee2018alma,bjerkeli2019kinematics,maret2020searching,oya2020substructures}.
%\vspace{-0.5cm}

Figure 8 illustrates the relative disc abundance for Kolmogorov-type subsonic turbulence models (M1b$-$M4b) and Burgers type of supersonic turbulence models (M5b$-$M8b). It appears that an increasing initial thermal state (8 K$-$14 K)) leads to more resolvable disc structures around the protostars, especially for cores that began collapsing with subsonic velocity dispersion. However, we caution readers as this apparent result  may be an artefact since the number of resolvable discs is influenced by the SPH particle resolution in our simulations. It is therefore important to acknowledge the role of the numerical resolution before reaching any final conclusion, in this case concerning the relationship between the number of resolvable discs and the initial thermal state of the gas cores. Also, for the initial thermal state and types of turbulence in our models, we do not find any correlation between the frequency of the final number of protostars and the protostars with resolvable disc structure.

\subsection{Disc radial profile}
As mentioned already, the effects of different turbulent realisations in terms of different spectral slopes corresponding to Kolmogorov and Burgers-type turbulence in our model sets M1a$-$M8a and M1b$-$M8b are apparent in the morphology of the collapsed gas cores. This subsequently has yielded different results in the form of the properties associated with the CDs produced in our calculations. For the second seed i.e. M1b$-$M8b, we find that a significant number of computationally resolvable discs are produced, and so we focus on this case in the remainder of this section. 
%It is important to mention here that selecting only self-gravitating discs to present radial profiles of the systems may ignore a companion (primary/secondary) from the binary system which fails to qualify as a protostar with a self-gravitating circumprimary/circumsecondary disc structure. Hence, if two sinks are marked as binary components (with a solid line in the radial profiles) then it may not always indicate the two companions of the same binary system in the model. 
In figures 9-13, we present disc radial profiles where the top four panels in Figure 9$-$13 are for Kolmogorov-type subsonic turbulence models while the bottom four panels show the Burgers type of supersonic turbulence models. Also, with a temperature step of 2 K, the initial thermal state in each model set i.e. M1b$-$M4b and M5b$-$M8b, is varied from 8 K$-$14 K (see also Table 1).
\vspace{-0.5cm}
\subsubsection{Surface density profile}
Figure 9 shows the radial profiles of the azimuthally-averaged surface density ($\Sigma_{\rm disc}$) of each CD. The radial profiles of the surface density can be divided into two main parts. In the first part, there exists a plateau of ($\Sigma_{\rm disc}$), and in the second part the surface density steeply declines. 

In the case of (subsonic Kolmogorov regime), in model M1b ($T_{\rm i}$ = 8 K), CDs associated with sinks 1, 2, and 8 indicate a dense inner region. For sink 1 (a binary companion) and 2 (isolated), the thick disc structure extends up to $\sim$ 12 au. Beyond it, the plateau converts into a sharp decline until the edge of the two discs. For disc around sink 8 (isolated object), the high surface density with a plateau last only up to $\sim$ 3 au, and beyond it a steep decline occurs making the outer part a thin disc structure.  In model M2b ($T_{\rm i}$ = 10 K), sink 7 and sink 11 (binary companions) show relatively dense inner discs when compared with sink 1 and sink 6 (isolated objects). For discs related to binary companions, the plateau in $\Sigma_{\rm disc}$ lasts only up to $\sim$ 3 au. However, the plateau $\Sigma_{\rm disc}$ in the isolated CDs extends to radii of $\sim$ 15 au. Beyond this, in both cases, a steep decline in the surface density is observed which continues until the edge of the respective discs. The CDs in model M3b ($T_{\rm i}$ = 12 K) indicate almost identical inner surface density structure and in most of the cases the plateau in $\Sigma_{\rm disc}$ lasts up to < 5 au. Beyond this, the disc surface density shows a decline until the edge of the respective CDs. Similarly, model M4b ($T_{\rm i}$ = 14 K) also shows nearly the identical surface density profiles for all the sinks.

In the case of (supersonic Burgers regime), in model M5b ($T_{\rm i}$ = 8 K), the discs around sinks 1, 2 (binary companions) exhibit an identical range of plateau in $\Sigma_{\rm disc}$ up to $\sim$ 10 au. The disc around sink 3 (isolated objects), however,  has a more extended plateau up to $\sim$ 20 au. The extended disc structure of sink 3 shows a sharp decline at the outer edge of the disc. In model M6b ($T_{\rm i}$ = 10 K), the two discs around sinks 3 and sink 4 (isolated objects) exhibit a smaller range of plateau in $\Sigma_{\rm disc}$ up to $\sim$ 6 au. Afterwards, both discs show a steep decline in the surface density until the edge of the respective discs. 
In model M7b ($T_{\rm i}$ = 12 K), the two discs around sinks 1 and sink 2 (binary companions) have  thick inner disc structures. For sink 1, the plateau in $\Sigma_{\rm disc}$ for the CD lasts up to $\sim$ 12 au while for sink 2 it lasts only up to $\sim$ 5 au. Afterward in both cases, the disc shows a steep decline in the surface density until the edge of the respective discs. The CDs associated with sinks 6 (isolated objects) has relatively thin inner disc structures. Also, the plateau in $\Sigma_{\rm disc}$ for CD in this case lasts up to $\sim$ 4 au. Beyond this, the disc shows a steep decline in their surface density until the edge of the disc. Model M8b ($T_{\rm i}$ = 14 K) has two discs associated with sink 1 and 4 (binary companion) and sink 7 (isolated object). The discs around the binary companion and the isolated object have a plateau in $\Sigma_{\rm disc}$ up to $\sim$ 12 au and  $\sim$ 20 au, respectively. %The outer part of the disc becomes thinner more in the case of sink 7 than in sink 4.
We do not find any significant influence of the nature of turbulence in the gas on the surface density profiles of the resulting CDs in all of our models.
\vspace{-0.5cm}
\subsubsection{Radial and azimuthal velocities profiles }
We now analyse the velocity structure within the CDs. Figures 10 and 11 depict the radial and azimuthal velocity structures within the discs. In the case of (subsonic Kolmogorov regime), in model M1b ($T_{\rm i}$ = 8 K), sink 1 (binary companion) forms a circumprimary disc.  The other two sinks are sink 2 and 8 (isolated objects) in this model. The inner circumprimary disc < 2 au has a mix of both radial and azimuthal velocity flows with no dominant velocity structure over another but in the outer part from $\sim$ 2 au until the edge of the disc at $\sim$ 15 au the azimuthal velocity continues to decline. At the edge of the disc, there are signs of an increase in the inward radial flow. The other interesting CD is associated with an isolated sink 2 in which a steady rotational flow and a declining inward radial flow are observed until $\sim$ 2.5 au. Beyond this, a significant decrease in rotational motion accompanied by a steady but inward radial flow is seen until the edge of the disc at $\sim$ 15 au. The CD associated with sink 8 in this model exhibits almost identical behaviour with no variations in small inward flow and nearly zero rotational flow within the entire disc structures. In model M2b ($T_{\rm i}$ = 10 K), sink 7 and sink 11 constitute the binary system. Interestingly, a mix of both inward radial and rotational motion is present within the inner part at $\sim$ 2 au of the circumprimary and circumsecondary. This is then followed by a weaker decline in both rotational motion and the inward radial motion of the circumsecondary disc than the circumprimary disc until the edge of the two disc structures at $\sim$ 7 au and $\sim$ 10 au, respectively. These two discs experience an infall of gas with an almost similar magnitude of the specific AM which is depositing material onto the discs via think spiral arms (see the top-right panel of Figures 4 and also Figure 6). Yet the more active radial velocity profile of the circumsecondary disc may hint at a possible mass sharing between the binary components where the smaller circumprimary disc continues to feed the more extended circumsecondary disc with a mechanism which could be based on intense accretion burst (see e.g. Figure 4, top panel \citep{riaz2021turbulence}). Also, it has been suggested that the net torque on the binary is positive for mass ratios close to unity, and that accretion always
drives the binary toward equal mass \citep{duffell2020circumbinary}. Model M3b ($T_{\rm i}$ = 12 K) has binary components (sink 6 and sink 17) but these do not form mutually the single binary system. Because these two sinks individually form a binary system with sink 9. The discs of sink 6 and sink 17 exhibit identical radial and azimuthal velocity profiles in terms of the fluctuating rotational and inward radial flows at the inner part of the disc which is followed by a gradual decrease up to the edge of the two discs. The only difference is that the former extends up to $\sim$ 30 au while the latter has a radius of $\sim$ 9 au. All isolated CDs around the remaining sinks in this model show nearly identical velocity structures in form of the in-ward radial and rotational motions which exhibit a smooth trend with nearly no fluctuations. In model M4b ($T_{\rm i}$ = 14 K), sinks 8 and sink 17 constitute a binary system and have circumprimary and circumsecondary discs, respectively. Inside the inner radius of $\sim$ 4 au, the circumprimary disc shows an small inward radial flow than the circumsecondary disc. The rotational flow within the inner radius for the former is significantly large than for the latter. Beyond this inner radius until the edge of the discs, the two discs show nearly identical decreasing inward radial flows. However, the rotational motion in the circumprimary disc remains stronger than in the circumsecondary disc. Amongst the isolated CDs, sink 1 and sink 2 show interesting behaviour. For sink 1, there exists a strong rotational flow and weak inward radial flow within the inner part of $\sim$ 2 au of the CD. This then declines in the outer part of the disc until its edge at $\sim$ 15 au. For sink 2, the inner part of the disc $\sim$ 2 au sustains a high rotational gas motion while a continuous decline is present in the inward radial velocity flow. The two types of flow at the outer edge of the disc are sustained. The remaining isolated CDs show nearly identical trends with little variations in rotational and inward radial flows. 

In the case of (supersonic Burgers regime), model M5b ($T_{\rm i}$ = 8 K) yields one isolated CD around sink 3, which is more massive and more extended when compared with the disc around sink 1 and 2 that constitute a binary system. Within the inner part of the disc around sink 3 up to $\sim$ 3 au, the radial velocity structure is identical with signs of inward flow. This behaviour changes to nearly zero inward motion at the outer part of the disc. Interestingly, a strong rotational flow within the inner dimension of the disc of sink 3 is evident, which declines significantly towards the edge of the disc. The CDs around sink 1 and 2 (binary system) more a less exhibit identical trend in terms of the radial velocity. However, the azimurthal velocity profiles in the discs of the two binary companions differ in terms of the circumprimary disc showing more rotational flow within the inner part of the disc than in case of the circumsecondary disc. Both discs show a decline in the rotational flow towards theie outer parts. In model M6b ($T_{\rm i}$ = 10 K), a somewhat similar radial velocity profile is seen for sink 3 and sink 4. However, as compared to model M5b, in model M6b there exist signs of inward radial flow at the edge of the two isolated CDs. A much stronger rotational flow is seen for disc associated with sink 4 than with sink 3. We suspect that the nearby binary system with its spiral arms (which are already exposed to higher specific AM gas infall) provides higher specific AM to the infalling gas onto the disc of sink 4 (see Figure 5 top-right panel and also Figure 6). Nonetheless, the rotational flow becomes weaker in the outer part of the two discs and approaches nearly a zero rotational flow at their respective outer edges of $\sim$ 40 au and $\sim$ 16 au, respectively. In model M7b ($T_{\rm i}$ = 12 K), a binary system with circumprimary and a circumsecondary discs are formed around sink 1 and 2. The outer part of the disc around primary companion i.e. sink 1 exhibits more inward gas flow than what is observed in the disc around secondary companion i.e. sink 2. The two discs of the binary companions show almost identical trend in their rotational gas motion where circumsecondary exhibits more strong rotation of material than circumprimary. Sink 6 (isolated object) shows more radial flow than the rotational flow within the disc. In model M8b ($T_{\rm i}$ = 14 K), yields a binary system constituted by sink 1 and 4. Both circumprimary and circumsecondary discs indicate presence of strong radial flow at the outer part of the disc. The inner part of the discs remain under the influence of strong rotational flow of the disc material. Sink 7 (isolated object) forms a disc with a steady radial and azimuthal flows within the disc structure.      
In general, we do not find any influence of the initial thermal state and the type of turbulence model that we explored (i.e. Kolmogorov and Burgers type of turbulence) in the parent gas core in models M1b$-$M8b on the CDs these models produce.     
\vspace{-0.5cm}

%Thermal structure (sound speed) in the disc governs the resistance of the gas to gravitational instability and hence the phenomenon of planet-formation as well as setting the chemical composition of the planet-forming material \citep{van2020temperature}. It is therefore important to investigate the young disc and its possible gravitationally unstable state. The disc thermal structure can also contribute toward luminosity outbursts (see e.g. \citealt{vorobyov2009variable}). The thermal state of the disc is also related to the dissipation of viscous energy via turbulence (i.e. turbulent viscosity) \citep{shakura1973black, lynden1974evolution, pringle1981accretion, nomura2002structure}. 

%To compute Mach numbers $\mathcal{M}_{\rm disc}$ and the temperature $T_{\rm disc}$, we use the following relations:
%To compute Mach numbers $\mathcal{M}_{\rm disc}$ inside the disc, we use the following relation:
%\begin{equation} \label{alpha_turb}
%\mathcal{M}_{\rm disc}=\frac{|v_{\rm \phi}|}{c_{\rm s}},
%\end{equation}
%and
%\begin{equation} \label{alpha_turb}
%T_{\rm disc}=\frac{c_{\rm s}^{2} \mu m_{\rm H}}{\gamma k_{\rm B}},
%\end{equation}
%where $v_{\rm \phi}$ and $c_{\rm s}$ are the azimuthal velocity and the sound speed in the circumstellar disc.  
\subsubsection{Keplerian velocity in the disc}
Keplerian rotation is a tell-tale sign of true, rotationally supported discs and allows sufficient time for gravitational instabilities, collisions, and accretion processes to facilitate the formation of planetary systems \citep{segura2018vla}. Observations show that large (r > 50 au) Keplerian discs are rare in Class 0 protostars \citep{maret2020searching}. In Figure 11, we explore the Keplerian velocity profile in the discs resulting from our simulation models (M1b$-$M8b). The presence of circumstellar discs around protostars and their connection to the growth of stellar mass is a fundamental aspect of star formation \citep{vorobyov2015variable,beltran2016accretion}.
Our simulation results regarding non-perfectly Keplerian flows within these circumstellar discs align with the idea that if the flows were entirely Keplerian, there might not be any net accretion onto the protostar. Inside the disc structure there is a centripetal force due to the gravity of the protostar, the centrifugal force that is generated by the rotational motion of the disc, and the pressure gradient within the disc. The deviations from strict Keplerian motion suggest that pressure gradients play a relevant role within these discs, contributing to the accretion process onto the protostars (see \citep{riaz2021turbulence}).
Faster rotation of disc material from larger to smaller radii as can be seen in Figure 12 is in line with theoretical expectations from accretion disc physics. This increase in the rotational velocity as material moves closer to the center reflects the gravitational pull exerted by the central protostar.
The comparison of sub-Keplerian velocities within the disc to radial flow (as indicated in Figure 10) is interesting. It indicates that the velocities associated with sub-Keplerian motion surpass the radial flow velocities, emphasizing the dominance of rotational forces within the disc structure and hence affecting mass accretion onto the protostar via disc. 

In the(subsonic Kolmogorov regime), in model M1b ($T_{\rm i}$ = 8 K), sink 2 and sink 8 (isolated objects) form circumstellar discs. The disc around sink 2 exhibits sub-Keplerian motion while a completey different behaviour is observed in the disc around sink 8. Sink 1 which is a circumprimary disc in a binary system also shows sub-Keplerian motion.  In model M2b ($T_{\rm i}$ = 10 K), the binary system constituted by sink 7 and sink 11 shows identical sub-Keplerian motion as indicated by the isolated disc systems in sink 1 and sink 6. In model M3b ($T_{\rm i}$ = 12 K), sink 6, sink 9 and sink 17, which are part of the two different binary systems, show the presence of the sub-Keplerian motion in the disc structures. Similarly, amongst the isolated discs, sink 2, 11, 16, and 24 show sub-Keplerian discs. Model M4b ($T_{\rm i}$ = 14 K), and the sinks therein similarly exhibit sub-Keplerian motion as observed in model M3b. In the supersonic Burgers regime, model M5b ($T_{\rm i}$ = 8 K) yields one isolated CD around sink 3 and a binary system formed by sink 1 and 2. The disc around sink 3 (an isolated system) 
 shows a weaker Keplerian motion as compared to the disc involved in the binary system. Model M6b ($T_{\rm i}$ = 10 K), also has two disc structures around sink 3 and sink 4 with sub-Keplerian motion. Model M7b ($T_{\rm i}$ = 12 K) has a binary system constituted by sink 1 and sink 2, which shows sub-Keplerian motion for the associated disc structures.  Sink 6 (isolated objects) exhibits weak Keplerian motion in its disc structure. In model M8b ($T_{\rm i}$ = 14 K), the binary system, which is constituted by sink 1 and 4 shows sub-Keplerian discs while sink 7 (isolated objects) exhibits relatively weaker sub-Keplerian motion when compared with the sub-Keplerian motion observed in the binary system.
%\vspace{-0.5cm}

\subsubsection{q-parameter profile}

In the previous section, we focused on turbulence in the CDs which can contribute to heating the disc via viscous dissipation hence influencing the tendency of the disc to fragment. However, the disc fragmentation also depends on the surface density. Therefore a more complete picture is provided by the Safronov$-$Toomre criterion. The parameter called q determines if a disc is prone to a gravitational instability (see \citealt{toomre1964gravitational}): 
\begin{equation} \label{transition}
q = \frac{c_{\rm s} \kappa}{\pi G \Sigma} < 1. 
\end{equation}
Here, $c_{\rm s}$ is the sound speed, $G$ is the gravitational constant, $\Sigma$ is the mass per unit surface area, and $\kappa$ is the epicyclic
frequency. 
%\begin{equation} \label{Jeans Radius}
%\kappa=\sqrt{\frac{GM}{r_{\rm disc}^{3}}}
%\end{equation}
%For a marginally unstable self-gravitating disc i.e $q$ = 1, the pressure gradient to the velocity field in the disc is expected to be sub-dominant compared to the disc self-gravity \citep{longarini2021investigating}. We believe that the criterion we follow in this work to define a self-gravitating CD (i.e. $M_{\rm disc}$/$M_{\rm star}$ $\geq$ $10^{-2}$), the effects of pressure gradients in CDs form in our models are weaker and hence can be ignored.

Figure 13 illustrates the $q$-parameter profile of the discs that are formed around the protostars in our simulation models M1b$-$M8b. In the case of (subsonic Kolmogorov regime), in model M1b ($T_{\rm i}$ = 8 K), the disc around sink 1 belongs to binary system. It shows an unstable circumprimary disc at a radius $\sim$ 10 au. We notice that for the circumprimary disc both $\Sigma_{\rm disc}$ as well as $\mathcal{M}_{\rm disc}$ are high at $\sim$ 10 au (see also Figures 7 and 10). The former decreases the $q$ while the latter being also high indicates lower sound speed (see equation 6) at a radius of $\sim$ 10 au and hence promotes $q$ to become < 1. Amongst the isolated CDs, sink 2 also becomes gravitationally unstable at $\sim$ 10 au. This disc shows high $\Sigma_{\rm disc}$ as well as high $\mathcal{M}_{\rm disc}$ at $\sim$ 10 au (see also Figures 7 and 10) and hence drives $q$ to become < 1. The disc around sink 8 exhibits stablity against gravitational fragmentation as $q$ remains > 1. In model M2b ($T_{\rm i}$ = 10 K), both companions (sink 7 and sink 11) show a marginally unstable disc at radii $\sim$ 5 au and $\sim$ 9 au, respectively. Both discs in the binary system show high enough $\Sigma_{\rm disc}$ and $\mathcal{M}_{\rm disc}$ at respective radii hence explaining their GI. The isolated discs around sink 1 and sink 6 have a wider disc dimension between 4 au and 17 au within which the GI prevails. These disc structures also show a much higher $\Sigma_{\rm disc}$ and $\mathcal{M}_{\rm disc}$ within the relevant disc dimension. This explains their tendency to fragment. In model M3b ($T_{\rm i}$ = 12 K), other than the discs around sink 6, sink 17, and sink 24, every other disc structure exhibits a tendency to become gravitationally unstable between radii of $\sim$ 6 au and 30 au. We find similar reasons for these gravitationally unstable discs when Figures 7 and 10 are also taken into account. We recommend the reader to see Figures 7 and 10 as these explain the reason of GI to prevail in these disc structures.

In the case of (supersonic Burgers regime), model M5b ($T_{\rm i}$ = 8 K) has one isolated CD around sink 3. The disc around sink 3 is unstable and can fragment within radii of  $\sim$ 8 au. This disc has high $\Sigma_{\rm disc}$ and $\mathcal{M}_{\rm disc}$ within the relevant range of disc radii and hence is subject to $q$ < 1. Sink 1 and 2 constitute a binary system in which circumprimary disc around sink 1 remains stable against gravitational instability (GI), while circumsecondary disc around sink 2 indicates possibility of gravitational fragmentation at a radius of $\sim$ 12 au. In model M6b ($T_{\rm i}$ = 10 K), the CD around sink 3 is under a strong influence of GI. The $q$ parameter becomes << 1 at the middle $\sim$ 6 au and at the edge $\sim$ 20 au of the disc. The disc around sink 3 has high $\Sigma_{\rm disc}$ but nearly transonic gas behaviour within the relevant disc dimension. We suspect that $\Sigma_{\rm disc}$ plays a more dominant role than the sound speed in the disc to drive the disc structure becoming unstable against GI. In contrast, the disc around sink 4 is largely stable against fragmentation except at a radius $\sim$ 10 au. Both $\Sigma_{\rm disc}$ and $\mathcal{M}_{\rm disc}$ for sink 4 explains the evolution of $q$-prameter. In model M7b ($T_{\rm i}$ = 12 K), the binary system is constituted by sink 1 and sink 2. The respective circumprimary and circumsecondary discs show a different $q$-prameter as the former is unstable against fragmentation between radii of $\sim$ 5 au and  $\sim$ 12 au while the latter is stable and shows no tendency to fragment throughout the disc dimension. The remaining sink 6 in this model which forms isolated CD is in gravitationally unstable state at $\sim$ 10 au. Model M8b ($T_{\rm i}$ = 14 K) has one binary system constituted by sink 1 and 4. The former is a circumprimary disc that remains stable against GI, while the latter that is a circumsecondary disc shows tendency of gravitational fragmentation at a radius of $\sim$ 15 au. The isolated CD around sink 7 does not show any signs of being unstable against disc fragmentation. 

%A typical evolution of the disc is related to emergence
%and decay of the spiral arms. The disc swings between the two states, the expanding state with spiral arms
%and the relatively circular, flat state. The transition timescale corresponds to the disc dynamical timescale \citep{tomida2017grand}. We quantify the time scale between the switch of the two states as follows:

%\begin{equation} \label{Jeans Radius}
%t_{\rm rot}=\sqrt{\frac{4 \pi ^{2} r_{\rm disc}^{2} }{ G M_{\rm \ast}}}
%\end{equation}
\vspace{-0.5cm}
\subsection{Caveats}
We perform hydrodynamical simulations to investigate the properties of the CD around YSOs. However, the magnetic field plays a crucial role not only in limiting the number of protostars forming in a collapsing gas core (e.g. \citep{commerccon2011collapse, peters2011interplay, price2008effect}), but also in affecting the morphology of the CD and controlling the mass of the disc $M_{\rm disc}$, the radius of the disc $R_{\rm disc}$, and the disc$-$star mass ratio $M_{\rm disc}$/$M_{\rm star}$ \citep{matsumoto2017circumstellar}. Another relevant feature in MHD modelling is its ability to address magnetic braking catastrophe which can influence disc properties significantly \citep{seifried2012magnetic, seifried2012disc} and \citep{hennebelle2011collapse,hennebelle2009disk}. 
The other limitation of our numerical scheme is neglecting radiative feedback processes, which can affect the GI in the CDs \citep{stamatellos2015migration, klassen2016simulating, mercer2019numerical}.  
\vspace{-0.5cm}
\section{Conclusions and outlook}
Our models based on Kolmogorov-type turbulence (i.e. M1b$-$M4b) yield more massive discs than Burgers-type models (i.e. M5b$-$M8b). On the lower end of the disc mass-spectrum less massive discs are produced in the former type of models than in the latter. Moreover, we find in the star-forming gas a wider range of CD masses in the case of Kolmogorov turbulence but not for Burgers type of turbulence.

For Kolmogorov-type and Burgers-type turbulence, the ranges of disc mass are 30.37 $M_{\rm Jup}$ $-$ 0.92 M$_{\odot}$, and 2.09 $M_{\rm Jup}$ $-$ 0.13 M$_{\odot}$, respectively.

Also, our results suggest that the ratio $M_{\rm disc}$/$M_{\rm star}$ is higher in models of Kolmogorov-type turbulence  than in models of Burgers-type turbulence. In the former case the ratio varies within the range of 0.029 to 0.920 M$_{\odot}$, while in the latter type the ratio varies between the range of 0.002 to 0.139 M$_{\odot}$.

In the case of molecular cores with Kolmogorov-type turbulence, discs with even smaller radii are formed than in molecular cores with Burgers-type turbulence. Also, a significant number of discs are formed with a radius of $\sim$ 15 au in both cases of Kolmogorov and Burgers types of turbulence in star-forming gas. However, a greater number of extended discs with radii greater than 15 au is yielded by the former type of turbulent models than the latter.

%Nearly an identical range of disc radii 10 $-$ 30 au is found for Class 0 objects in our hydrodynamical simulation models. However, the mass range for Class 0 objects is obtained as 0.73 $M_{\rm Jup}$ $-$ 0.11 M$_{\odot}$, and 0.2 $M_{\rm Jup}$ $-$ 0.028 M$_{\odot}$ in Kolmogorov-type of subsonic turbulence (i.e. M1b$-$M4b) and in Burgers type of supersonic turbulence (i.e. M5b$-$M8b), respectively. 

%Misaligned CDs with respect to the axis of rotation of the parent gas core is a common feature found associated with Class 0 discs in our models. The degree of misalignment is found independent of both the initial thermal and turbulent characteristics of the parent gas core. Moreover, a wide range of possible degrees of misalignment  0$^{\circ}$ and 165$^{\circ}$ in CDS are found in our calculations.

We do not find any correlation between the disc radius and its mass over the explored range of initial thermal states and the type of turbulence of the collapsing gravoturbulent gas cores. 

The majority of the CDs in our simulations associated with isolated and binary companions are formed with a radius of $\sim$ 16 - 21 au. For systems of discs with radii in this range, the gas infall from the envelope onto the edge of the discs exhibits extreme variations in the specific AM.

We suggest that the radial profile of the CDs around class 0 objects is not influenced profoundly by the turbulent velocity field which is imposed as the initial conditions.

Small discs with a radius of less than 100 au are expected to be produced by the magnetized turbulent model \citep{matsumoto2017circumstellar}. We have demonstrated in our hydrodynamical simulation models that even an unmagnetised prestellar core can still produce a significant number of CDs with radii in the range of $\sim$ 7 au to $\sim$ 40 au via gravoturbulent collapse.

\citet{takaishi2021new} recently explored the formation pathway of a counter-rotating circumstellar disc in triple systems, which can explain the recently observed young counter-rotating discs by the Atacama Large Millimeter/submillimeter
Array (ALMA). Our model also produced such a system of counter-rotating discs whose origin lies in the intrinsic turbulence of the prestellar gas core that promotes non-uniform specific AM in the star-forming gas. We aim to focus and further investigate GI in such systems of counter-rotating Class 0 discs.

%Several MHD simulations have shown that a magnetic field can effectively remove the AM of collapsing gas via magnetic braking, and can limit the outer radii of Keplerian discs to within $\sim$ 10 au (see for example \citep{allen2003collapse, mellon2008magnetic, lee2011rotating, dapp2012bridging, yen2015no, segura2016vla}). We, aim to perform our calculations with MHD models to investigate the additional effects of magnetic field on the formation of small-scale self-gravitating discs which we report in this work. 
\vspace{-0.5cm}
\section*{Acknowledgements}

This research was partially supported by the supercomputing infrastructure of the NLHPC (ECM$-$02). The authors acknowledge the Kultrun Astronomy Hybrid Cluster (projects ANID Programa de Astronomia Fondo Quimal QUIMAL 170001, ANID PIA ACT172033, and Fondecyt Iniciacion 11170268) for providing HPC resources that have contributed to the research results reported in this paper. Also, the Geryon cluster at the Centro de Astro-Ingenieria UC was partially used for the calculations performed in this paper. BASAL CATA PFB-06, the Anillo ACT-86, FONDEQUIP AIC-57, and QUIMAL 130008 provided funding for several improvements to the Geryon cluster. 

RR remains thankful for funding through Agencia Nacional de Investigaci\'on y Desarrollo (ANID) (project code SA77210037).

DRGS gratefully acknowledges support by the ANID BASAL projects ACE210002 and FB210003, as well as via the Millenium Nucleus NCN19-058 (TITANs). DRGS thanks for funding via Fondecyt Regular (project code 1201280).

SV wishes to thank Prof. Dr. R. Keppens and Prof.  Dr. S. Poedts for providing access to the KUL supercomputing cluster Thinking while initially developing and testing the code that was used in this work. He also gratefully acknowledges the support of the KUL HPC team and the NLHPC (ECM$-$02) which provides continuous support while maintaining and developing the codes used in this paper. 

RSK acknowledges financial support from the Heidelberg cluster of excellence EXC 2181 (Project-ID 390900948) ``STRUCTURES'' funded by the German Excellence Strategy, and from the European Research Council in the ERC Synergy Grant ``ECOGAL'' (grant 855130), and from the German Ministry for Economic Affairs and Climate Action in project ``MAINN'' (funding ID 50OO2206). He also thanks for computing resources provided by {\em The L\"{a}nd} through bwHPC and DFG through grant INST 35/1134-1 FUGG and for data storage at SDS@hd through grant INST 35/1314-1 FUGG.

We are grateful to the anonymous referee who has provided a number of comments and suggestions that have greatly improved the content of the manuscript.

%%%%%%%%%%%%%%%%%%%%%%%%%%%%%%%%%%%%%%%%%%%%%%%%%%
%\section*{Data Availability}

%The inclusion of a Data Availability Statement is a requirement for articles published in MNRAS. Data Availability Statements provide a standardised format for readers to understand the availability of data underlying the research results described in the article. The statement may refer to original data generated in the course of the study or to third-party data analysed in the article. The statement should describe and provide means of access, where possible, by linking to the data or providing the required accession numbers for the relevant databases or DOIs.
\vspace{-0.5cm}
\section*{Data Availability}
The data underlying this article will be shared on reasonable request to the corresponding author.
%%%%%%%%%%%%%%%%%%%% REFERENCES %%%%%%%%%%%%%%%%%%
% The best way to enter references is to use BibTeX:

\bibliographystyle{mnras}
\vspace{-0.5cm}
\bibliography{example} % if your bibtex file is called example.bib
% Alternatively you could enter them by hand, like this:
% This method is tedious and prone to error if you have lots of references
%\begin{thebibliography}{99}
%\bibitem[\protect\citeauthoryear{Author}{2012}]{Author2012}
%Author A.~N., 2013, Journal of Improbable Astronomy, 1, 1
%\bibitem[\protect\citeauthoryear{Others}{2013}]{Others2013}
%Others S., 2012, Journal of Interesting Stuff, 17, 198
%\end{thebibliography}

%%%%%%%%%%%%%%%%%%%%%%%%%%%%%%%%%%%%%%%%%%%%%%%%%%

%%%%%%%%%%%%%%%%% APPENDICES %%%%%%%%%%%%%%%%%%%%%
\vspace{-0.5cm}
\appendix

\section{Resolution study}
We provide a numerical resolution study to probe the stability and merit of the analysis of the circumstellar disc structures formed in our simulations. To investigate the evolution of the circumstellar discs, we select initial conditions identical to our model M2b. We perform three additional simulations for model M2b with SPH particle number $N_{\rm P}$ = 2000, 20000, and 100000. Moreover, we employ the same method to characterise the resulting disc properties as in the cases of our calculations in the main part of this paper. Criterion related to the sink formation also remain identical to our main calculations presented in this work. We terminate these additional calculations when each resolution model reaches $\xi$ = 15 \%. These additional models have yielded various numbers of sink particles (protostars) during model evolution. We select the most massive sink particle in each of these models and investigate the associated disc structure around it. Figure A1 illustrates the evolution of the disc mass with time in these three additional calculations of the collapsing prestellar gas cores. In the case of a low-resolution model with $N_{\rm P}$ = 2000, the collapsing gas takes longer time to create the sink particle which appears in the simulation at $t$ = $4.35 \times 10^{4}$~ yr and the resulting most massive sink particle attains the peak mass of 0.16 M$_{\odot}$ (64 SPH particles) at $\sim$ $t$ = $4.5 \times 10^{4}$~ yr. Immediately after this, the poorly resolved circumstellar disc quickly evolved viscously and at $\sim$ $t$ = $5.0 \times 10^{4}$~ yr the circumstellar discs started losing mass and the disc almost disappeared.
In the case of a medium-resolution model with $N_{\rm P}$ = 20000, the collapsing gas creates the sink particle at $t$ = $3.80 \times 10^{4}$~ yr and the resulting most massive sink particle gradually accumulates the surrounding material to form its circumstellar disc. The disc attains a peak mass of 0.075 M$_{\odot}$ (300 SPH particles) at $\sim$ $t$ = $4.2 \times 10^{4}$~ yr. After this, the circumstellar disc maintains the disc structure and continues to evolve over time until we finish the calculations at $t$ = $4.5 \times 10^{4}$~ yr. The highest resolution case amongst these additional three calculations, with $N_{\rm P}$ = 100000 (which is still a low resolution case when compared with our main calculations performed with $N_{\rm P}$ = 250025), creates sink particle at $t$ = $3.75 \times 10^{4}$~ yr. A sharp increase in the mass accumulation within the circumstellar disc around the most massive sink particle is observed in this model. The disc attains its peak mass of 0.16 M$_{\odot}$ (3200 SPH particles) at $\sim$ $t$ = $4.1 \times 10^{4}$~ yr. After this, the disc mass initially declines but soon the disc recovers and maintains its mass accumulation, which later also shows a rise at the time $t$ = $4.4 \times 10^{4}$~ yr when we terminate the model evolution.

In these three cases of resolution study performed with SPH particle number $N_{\rm P}$ = 2000, 20000, and 100000, the respective mass resolutions are $2.5 \times 10^{-1}$ M$_{\odot}$, $2.5 \times 10^{-2}$ M$_{\odot}$, $5.0 \times 10^{-3}$ M$_{\odot}$, respectively. In the lowest resolution case, the disc mass during the entire model evolution is not resolvable. In the remaining two cases of resolution, the disc mass of the most massive sink particle is well inside the range of mass resolution. In the main calculations, which is performed using $N_{\rm P}$ = 250025 with a minimum mass resolution of $1.999 \times 10^{-3}$ M$_{\odot}$, the disc structures presented in this work satisfy the minimum mass resolution criterion.     
\begin{figure}
	 %To include a figure from a file named example.*
	% Allowable file formats are eps or ps if compiling using latex
	% or pdf, png, jpg if compiling using pdflatex
	\includegraphics[angle=0,scale=0.525]{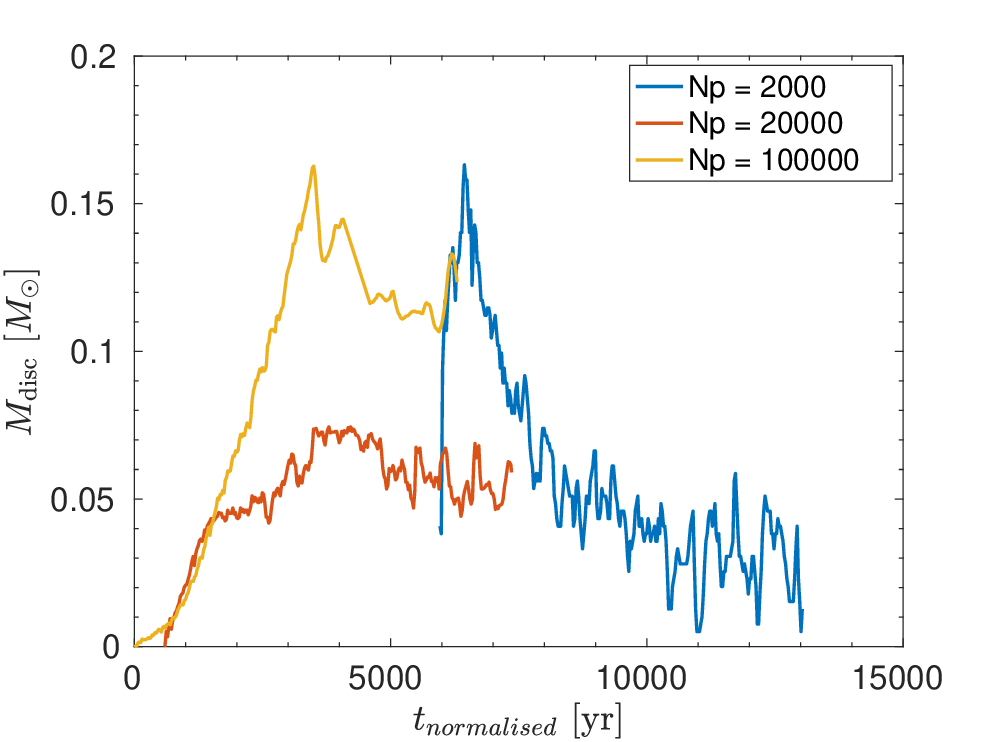}
	\caption{The evolution of the disc with time (normalised) in calculations of the collapse of model M2b (as in the main calculation) but performed with three different numerical
resolutions: $N_{\rm P}$ = 2000, 20000, 100000 SPH particles. The mass of a single SPH particle, the peak mass of the disc, and the SPH particle number constituting the disc in these three resolution models are ($2.5 \times 10^{-3}$ M$_{\odot}$, $2.5 \times 10^{-4}$ M$_{\odot}$, and $5.0 \times 10^{-5}$ M$_{\odot}$), (0.160 M$_{\odot}$, 0.075 M$_{\odot}$, and 0.160 M$_{\odot}$), and (64, 300, 3200), respectively. The disc mass and the time in the figure are shown in units of solar mass and years, respectively.}
	\label{fig:figur6}
\end{figure}

\begin{table} \label{tbl-1}
%\begin{flushleft}
\centering
\caption{Summary of the models in set M1a$-$M4a with the second random seed (corresponding to two different realisations of turbulence, with the same statistical properties). The table is constructed for sink particles (protostars) that form around them a CD structure. All sinks produced in models are included in the table. The entire table is constructed at the points in time when SFE $\xi$ reaches 15 \% in each model. The table describes the identity of protostar (Sink), the mass of the protostar ($M_{\rm star}$), the mass of the disc ($M_{\rm disc}$), the radius of the disc ($R_{\rm disc}$), and the disc$-$star mass ratio ($M_{\rm disc}$/$M_{\rm star}$). The mass and the radius are provided in solar mass units and in au, respectively.       }
%\headline{$\xi$ = 2 \%}

%. $\parallel$. $\perp$ symbols for parallel and perpendicular
\label{tab:Table1}
\begin{tabular}{ccccc} % four columns, alignment for each

%\begin{tabular}{ccccc}
\multicolumn{5}{|c|}{Model M1a, $\xi$ = 15 \% }\\
\hline
\hline
Sink & $M_{\rm star}$ (M$_{\odot}$) & $M_{\rm disc}$ (M$_{\odot}$) & $R_{\rm disc}$ (au) & $M_{\rm disc}$/$M_{\rm star}$ \\
%\tableline
\hline
1   & 0.170  &   0.005    &   18.994 & 0.030  \\
8   & 0.211  &   0.007    &   15.647 & 0.036  \\
10  & 0.215  &   0.0007   &   18.353 & 0.003  \\
13  & 0.152  &   0.0009   &   11.948 & 0.005  \\
\hline
%\tableline
\multicolumn{5}{|c|}{Model M2a, $\xi$ = 15 \% }\\
\hline
\hline
Sink & $M_{\rm star}$ (M$_{\odot}$) & $M_{\rm disc}$ (M$_{\odot}$) & $R_{\rm disc}$ (au) & $M_{\rm disc}$/$M_{\rm star}$ \\
%\tableline
\hline
1   & 0.175  &   0.0005   &   5.976  & 0.002  \\
4   & 0.189  &   0.0002   &   9.425  & 0.001  \\
6   & 0.193  &   0.0001   &   4.719  & 0.0005  \\
9   & 0.180  &   0.0007   &   4.929  & 0.003  \\
13  & 0.113  &   0.0089   &   12.895 & 0.078  \\

\hline

\multicolumn{5}{|c|}{Model M3a, $\xi$ = 15 \% }\\
\hline
\hline
Sink & $M_{\rm star}$ (M$_{\odot}$) & $M_{\rm disc}$ (M$_{\odot}$) & $R_{\rm disc}$ (au) & $M_{\rm disc}$/$M_{\rm star}$ \\
%\tableline
\hline
1   & 0.158  &   0.0003  &   3.911   & 0.001 \\
5   & 0.216  &   0.0003  &   2.644   & 0.001 \\
12  & 0.065  &   0.0014  &   9.524   & 0.021 \\
\hline
%\tableline

\multicolumn{5}{|c|}{Model M4a, $\xi$ = 15 \% }\\
\hline
\hline
Sink & $M_{\rm star}$ (M$_{\odot}$) & $M_{\rm disc}$ (M$_{\odot}$) & $R_{\rm disc}$ (au) & $M_{\rm disc}$/$M_{\rm star}$ \\
%\tableline
\hline
1   & 0.205  &   0.0018   &   10.577  & 0.008  \\
6   & 0.121  &   0.0008   &   11.690  & 0.006  \\
10  & 0.141  &   0.0006   &   4.713   & 0.004  \\
26  & 0.206  &   0.0007   &   3.205   & 0.003  \\
%31  & 0.006  &   0.0002   &   4.633   & 0.033 & iso & $-$ & yes \\
%32  & 0.003  &   0.005    &   9.897   & 1.860 & iso & 25$^{\circ}$ & yes\\
\hline
%\tableline

\end{tabular}
%\end{flushleft}
\end{table}

\begin{table} \label{tbl-1}
%\begin{flushleft}
\centering
\caption{Summary of the models in set M5a$-$M8a with the second random seed (corresponding to two different realisations of turbulence, with the same statistical properties). The table is constructed for sink particles (protostars) that form around them a CD structure. All sinks produced in models are included in the table. The entire table is constructed at the points in time when SFE $\xi$ reaches 15 \% in each model. The table describes the identity of protostar (Sink), the mass of the protostar ($M_{\rm star}$), the mass of the disc ($M_{\rm disc}$), the radius of the disc ($R_{\rm disc}$), and the disc$-$star mass ratio ($M_{\rm disc}$/$M_{\rm star}$). The mass and the radius are provided in solar mass units and in au, respectively.   }
%\headline{$\xi$ = 2 \%}

%. $\parallel$. $\perp$ symbols for parallel and perpendicular

\begin{tabular}{ccccc}
\multicolumn{5}{|c|}{Model M5a, $\xi$ = 15 \% }\\
\hline
\hline
Sink & $M_{\rm star}$ (M$_{\odot}$) & $M_{\rm disc}$ (M$_{\odot}$) & $R_{\rm disc}$ (au) & $M_{\rm disc}$/$M_{\rm star}$ \\
%\tableline
\hline
1   & 0.288  &   0.0008   &   8.624   & 0.002  \\
3   & 0.197  &   0.002    &   8.989   & 0.010  \\
4   & 0.102  &   0.020    &   29.993  & 0.198  \\
5   & 0.106  &   0.003    &   9.597   & 0.036  \\
6   & 0.029  &   0.0008   &   17.583  & 0.027  \\
7   & 0.031  &   0.007    &   28.029  & 0.225  \\
\hline

\multicolumn{5}{|c|}{Model M6a, $\xi$ = 15 \% }\\
\hline
\hline
Sink & $M_{\rm star}$ (M$_{\odot}$) & $M_{\rm disc}$ (M$_{\odot}$) & $R_{\rm disc}$ (au) & $M_{\rm disc}$/$M_{\rm star}$ \\
%\tableline
\hline
1   & 0.290  &   0.025    &   38.662  & 0.088  \\
2   & 0.179  &   0.005    &   39.768  & 0.028  \\
3   & 0.119  &   0.002    &   39.943  & 0.016  \\
6   & 0.114  &   0.029    &   48.277  & 0.257  \\
7   & 0.057  &   0.0056   &   39.610  & 0.098  \\

\hline
\multicolumn{5}{|c|}{Model M7a, $\xi$ = 10 \% }\\
\hline
\hline
Sink & $M_{\rm star}$ (M$_{\odot}$) & $M_{\rm disc}$ (M$_{\odot}$) & $R_{\rm disc}$ (au) & $M_{\rm disc}$/$M_{\rm star}$ \\
%\tableline
\hline
1   & 0.397  &   0.015    &   33.756 & 0.040  \\
2   & 0.101  &   0.018    &   23.239 & 0.179  \\
\hline

\multicolumn{5}{|c|}{Model M8a, $\xi$ = 15 \% }\\
\hline
\hline
Sink & $M_{\rm star}$ (M$_{\odot}$) & $M_{\rm disc}$ (M$_{\odot}$) & $R_{\rm disc}$ (au) & $M_{\rm disc}$/$M_{\rm star}$ \\
%\tableline
\hline
1   & 0.309  &   0.0001    &   22.567 & 0.0003  \\
4   & 0.424  &   0.0002    &   12.153 & 0.0004  \\
\hline
\end{tabular}
%\end{flushleft}
\end{table}

\section{Morphology of seed 1 models}
In this section, we briefly discuss the morphology of the gravoturbulent core collapse related to models of seed 1 i.e. M1a$-$M8a. As discussed previously in this paper, these models yield a fewer number of computationally resolved  discs hence we do not focus our analysis related to circumstellar disc properties in the case of seed 1 (see Figure B1, and the relevant Tables A1 and A2). 
%However, the misalignment in the CDs is complementary to the observations in star-forming regions \citep{stapelfeldt2013hst, jensen2014misaligned, ohashi2022misaligned} 

Figure B1 has two major (left and right) panels each representing a set of models M1a$-$M4a and M5a$-$M8a, respectively. In the left major panel, model M1a (top-left) shows two binary systems constituted by the set of companions (sink 1, sink 2) and (sink 10, sink 13), respectively. The first binary system of sinks 1 and 2 indicates two massive circumprimary and circumsecondary discs each connected with spiral structures that feed the two discs via material infall from the surroundings. The other binary system comprised of sinks 10 and 13 exhibits less prominent discs around the binary companions. Nonetheless, this binary system is also well connected with the surrounding gas via thick spiral arms that supply material to the binary system. More importantly, in the reference xz-plane of the collapsed gas core, the visible spiral structures in both of the binary systems in this model are indicative of the extremely disoriented CDs evolving in the embedded phase. In model M2a (top-right panel), there exist two binary systems and an isolated system of CDs. The two binary systems are constituted by sinks 1 and 4, and sinks 6 and 9, respectively. Sink 13 forms an isolated CD in this model. The two binary systems and the one isolated system show active material flow from the surroundings to the respective CDs. The isolated CD indicates a massive gas reservoir with spiral arms feeding the CD more actively than any other CDs present in the model. Interestingly, in the reference xz-plane of the collapsed gas core, the visible spiral structures in both of the binary systems as well as in the isolated system in this model are indicative of the extremely disoriented CDs evolving in the collapsing gas core. In model M3a (bottom-left panel), there exist two binary systems and an isolated system of CDs. The two binary systems are constituted by sinks 1 and 5, and sinks 3 and 7, respectively. The isolated CD is formed around sink 12. Both binary systems show signs of spiral density waves supplying CDs the material from the surroundings. The first binary system is residing in a less dense reservoir of surrounding gas than the second binary system. The isolated CD also is connected with the stream of gas which also connects the two binary systems. Due to the random gas motion which is mainly caused by turbulence, the reference xz-plane of the collapsed gas core still indicates the visible spiral structures in both of the binary systems. This shows extreme disorientation of the CDs associated with binary companions. However, the isolated system in this model exhibits an edge-on view of the significantly misaligned isolated CD in this model. In model M4a (bottom-right panel), we have two binary systems constituted by sinks 1 and 6, and sinks 10 and 26, respectively. The other two sinks are sink 17 and sink 29 but these two do not form a CD around the protostars.  The two binary systems reside in dense gas reservoirs and hence are active in their respective disc growth over time. However, the first binary system exhibits an edge-on view and the second a face-on view with visible spiral arms. This extremely misaligned situation between the two binary systems evolving in the same collapsing gas core is indicative of multi-directional gas flows primarily due to the turbulence present in the gas. For the second set of models M5a$-$M8a (see the major-right panel and the four sub-panels within), the behaviour is somewhat similar and there exists a mix of orientations of the CDs associated with both binary and isolated systems. We also see the CDs forming in the gas cores with subsonic velocity dispersions are less spatially scattered than the CDs which are formed in the case of supersonic velocity dispersions. This also is true in model set M1b$-$M8b.  

\begin{figure}
	 %To include a figure from a file named example.*
	% Allowable file formats are eps or ps if compiling using latex
	% or pdf, png, jpg if compiling using pdflatex
 \includegraphics[width=\columnwidth]
% {vertical_B1.jpg}
{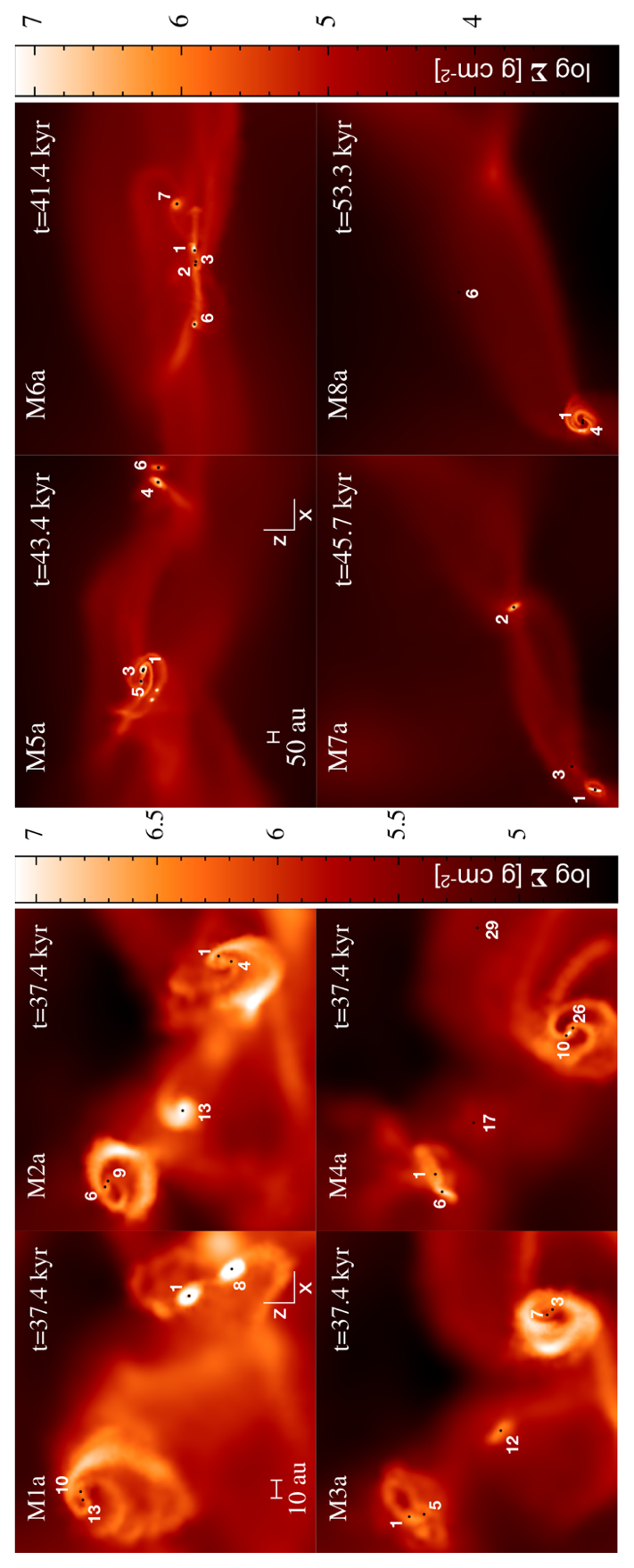}

	\caption{Morphology of the natal core collapse for models M1a$-$M4a and models M5a$-$M8a at the end of our computation when star formation efficiency (SFE) in each model reaches $\xi$ = 15 \%. Each panel (the xz-plane) shows in logarithmic scale the projected column density ($ \Sigma$) integrated along the y-axis in g cm$^{-2}$. Sink particles (protostars) are shown as black dots and are all numbered by following the order in which they are formed during the evolution of each model. The spatial scales depicted in the panels of models M1a and M5a differ and hold for M1a$-$M4a and M5a$-$M8a, respectively. Each panel for M1a$-$M4a and for M5a$-$M8a represents a box size of 170 x 180 au and 1250 x 1200 au, respectively. Colour in the online edition.}
	\label{fig:figur6}
\end{figure}
%\begin{figure}
	 %To include a figure from a file named example.*
	% Allowable file formats are eps or ps if compiling using latex
	% or pdf, png, jpg if compiling using pdflatex
%	\includegraphics[width=\columnwidth]{myradiusmass1.png}
%	\includegraphics[width=\columnwidth]{myradiusmass2.png}
%	\caption{Simulation results for models M1a$-$M4a (top panels) and M1b$-$M4b (bottom panels) at the end of our computation when star formation efficiency (SFE) in each model reaches $\xi$ = 2 \%. Each panel (the xy-plane) shows in logarithmic scale the projected column density ($ \Sigma$) integrated along the z-axis in g cm$^{-2}$. Sink particles (protostars) are shown as red dots. The arrows in each panel mark the primary (p) and secondary (s) components of MMPB located in the cluster. Colour in the online edition.}
%	\label{fig:figur6}
%\end{figure}
%If you want to present additional material which would interrupt the flow of the main paper,
%it can be placed in an Appendix which appears after the list of references.

%%%%%%%%%%%%%%%%%%%%%%%%%%%%%%%%%%%%%%%%%%%%%%%%%%

% Don't change these lines
\bsp	% typesetting comment
\label{lastpage}
\end{document}